\documentclass[11pt,a4paper]{article}
\pdfoutput=1
\usepackage{jheppub}
\usepackage{amsmath,epsfig}
\usepackage{mathrsfs}
\usepackage{amssymb}
\usepackage{mathtools}
\usepackage{graphics}
\usepackage{cancel}
\usepackage{slashed}
\usepackage{url}
\usepackage{hyperref}
\usepackage{color}
\usepackage[normalem]{ulem}
\usepackage{xspace}
\usepackage{xcolor}

\newcommand{\bremm}{bremsstrahlung\xspace}
\newcommand{\Bremm}{Bremsstrahlung\xspace}
\newcommand{\ie}{\textit{i.e.,~}}
\newcommand{\eg}{\textit{e.g.,~}}

\newcommand{\be}{\begin{equation}}
\newcommand{\ee}{\end{equation}}
\newcommand{\bea}{\begin{eqnarray}}
\newcommand{\eea}{\end{eqnarray}}

\preprint{IFIC/19-42}
\title{Searches for Atmospheric Long-Lived Particles }

\author[a]{C.~Arg\"uelles,}
\author[b]{P.~Coloma,}
\author[b]{ P.~Hern\'andez,}
\author[b]{V.~Mu\~noz,}

\affiliation[a]{Dept. of Physics, Massachusetts Institute of Technology, Cambridge, MA 02139, USA}
\affiliation[b]{Instituto de F\'{\i}sica Corpuscular, Universidad de Valencia and CSIC, 
 Edificio Institutos Investigaci\'on, Catedr\'atico Jos\'e Beltr\'an 2, 46980 Spain}

\abstract{Long-lived particles are predicted in  extensions of the Standard Model that involve relatively light but very weakly interacting sectors. In this paper we consider the possibility that some of these particles are produced in  atmospheric  cosmic ray showers, 
and their decay intercepted by neutrino detectors such as IceCube or Super-Kamiokande.  We present the methodology and evaluate the sensitivity of these searches in various scenarios,  including extensions  with heavy neutral leptons in models of massive neutrinos, models with an extra $U(1)$ gauge symmetry, and a combination of both in a $U(1)_{B-L}$ model. Our results are shown as a function of the production rate and the lifetime of the corresponding long-lived particles.}

\keywords{Beyond Standard Model, Neutrino physics, Long-lived Particles}

\begin{document}

\maketitle
\section{Introduction}

The Standard Model (SM) of particle physics has demonstrated to be an extremely predictive theory, which however cannot be complete. In particular, it lacks a mechanism to generate neutrino masses, dark matter, and the baryon asymmetry of the Universe that we observe today. While LHC upgrades and future colliders will keep pushing the energy frontier forward in order to search for new physics at high energies, it is important to keep in mind that the physics beyond the Standard Model (BSM) may very well involve weakly interacting particles at low scales instead, which in minimal extensions of the SM are expected to be relatively long-lived. In fact, the search for such long-lived particles (LLP) in colliders and beam-dump experiments has become a field of intense activity in recent years and several experiments have been proposed to conduct dedicated searches~\cite{Alekhin:2015byh,Feng:2017uoz,Curtin:2018mvb,Ariga:2018zuc}. 

In this work, we study the potential of atmospheric neutrino detectors to search for these particles since they can also be copiously produced in atmospheric showers. In order to do this, we compute the expected flux of exotic LLPs being produced in proton-nucleus collisions in the atmosphere, either through proton \bremm or as a product of meson decays. After traveling typical distances of tens of kilometers through the atmosphere, such LLPs may subsequently decay within the detector volume. 
The signal would therefore be an excess of events where the interaction vertex is contained in the fiducial volume of the detector. These are a priori indistinguishable from an atmospheric neutrino interactions, which therefore constitute an irreducible background to this search. 

The experimental setup considered in this work has two main advantages with respect to searches in laboratory experiments. First, the center-of-mass energy in proton-nucleus collisions in the atmosphere will typically be much higher than at beam-dump experiments, allowing to produce an abundant flux of heavy mesons and tau leptons. Secondly, the size of these huge detectors results in larger decay volumes and could be optimal in the searches for particles with long lifetimes, of about $\mathcal{O}(10)$~km in the laboratory frame. In this work, we will be considering two experimental setups: Super-Kamiokande, most sensitive to events with energies below 100~GeV; and Icecube, sensitive to events with energies at the TeV scale and above. Therefore, the results between the two will be largely complementary as they will be sensitive to different values of the lifetime of the particle in its rest frame, $\tau$. As we will see, the best sensitivities will be achieved for $\tau/m \sim \mathcal{O}(1)~\mathrm{km}/\mathrm{GeV}$ for Super-Kamiokande and $\tau/m \sim \mathcal{O}(10)~\mathrm{m}/\mathrm{GeV}$ for Icecube. 

In this work, for concreteness, we will focus on two minimal and theoretically well-motivated extensions of the SM: (1) a scenario with heavy neutral leptons (HNL), that could be responsible for generating neutrino masses \cite{Minkowski:1977sc,GellMann:1980vs,Yanagida:1979as,Mohapatra:1979ia} and possibly the baryon asymmetry of the Universe~\cite{Akhmedov:1998qx,Asaka:2005pn}, and (2) a model with an extra  $U(1)$ symmetry which, after being broken, leads to a massive dark photon and a portal to the dark sector~\cite{Galison:1983pa,Holdom:1985ag}. In the first case we focus on LLPs in the mass range between the Kaon and D-meson masses, where existing
laboratory bounds are weaker \cite{Atre:2009rg,Drewes:2015iva,Bryman:2019bjg,Bryman:2019ssi}, while in the latter we will consider dark photons below the proton mass.  The search for lighter, MeV range, atmospheric sterile neutrinos has been considered before in Super-Kamiokande \cite{Kusenko:2004qc,Asaka:2012hc} and IceCube \cite{Masip:2014xna}. Our analysis significantly improves the methodology of these earlier studies.

In the most minimal models with just HNL or dark photons, both production and decay of the LLP are highly correlated and, thus, longer lifetimes also mean lower production rates. However, in non-minimal extensions this may no longer be the case, as different mechanisms may be involved in production and decay processes. An example of this framework is provided by a $B-L$ model with HNL \cite{Batell:2016zod}: in this case, production of the HNL could take place through the $B-L$ gauge interaction \cite{Mohapatra:1980qe, Davidson:1978pm}, while the decay may occur via its mixing with the SM neutrinos. Such scenario will also be discussed in our work. Our results will be presented from a  model-independent perspective, considering a more general BSM scenario where the production and decay rates could be decoupled, and we will discuss the interpretation of our results in the three scenarios outlined above. 

The paper is organized as follows. In Section ~\ref{sec:production} we discuss the production of LLPs and describe the method to compute LLP fluxes 
from two main sources: meson decays (two- and-three body) and proton \bremm.
In Section~\ref{sec:detect} we present the general procedure to compute LLP signals in a neutrino detector which requires defining
 detector effective areas for decaying events, detector efficiencies, backgrounds and the data samples that will be used to extract limits on LLPs. In Section~\ref{sec:LLP} we focus on three BSM scenarios and use the results of the previous sections to compute the required fluxes. In sections ~\ref{sec:results} we present the sensitivities to LLPs of searches in IceCube and Super-Kamiokande,  and our conclusions are drawn in section \ref{sec:conclu}.

\section{LLP production in atmospheric showers}
\label{sec:production}

The flux of any SM particle produced in the atmosphere is usually given as a function of the slant depth, $X$, related to the integral of the density in the direction of the particle from the top of the atmosphere to the production point at distance $\ell$ \cite{Gondolo:1995fq}:
\begin{eqnarray}
X(\ell,\theta) = \int_\ell^{\ell_{\rm max} }\rho[h(l,\theta)] ~dl, 
\end{eqnarray}
where $\rho(h)$ is the atmospheric density at height $h$, and $\theta$ the zenith angle defined by the trajectory of the particle.
 The distance $\ell$ is related to the height $h$ and to the zenith angle $\theta$ as: 
\begin{eqnarray}
\label{eq:ell}
h(\ell, \theta) = \sqrt{R_\oplus^2 + 2 \ell R_\oplus \cos\theta + \ell^2} - R_\oplus,
\end{eqnarray}
where $R_\oplus \simeq 6370$km is the Earth radius. In our calculations, we consider $h(\ell_{\rm max}, \theta) =80$~km as the maximum vertical height of the atmosphere.

In our calculations, we have used the Matrix Cascade Equation (MCEq) Monte Carlo software~\cite{Fedynitch:2015zma,Fedynitch:2012fs} to compute the fluxes for the parent mesons and protons in the atmosphere, with the SYBILL-2.3 hadronic interaction model~\cite{Fedynitch:2018cbl}, the Hillas-Gaisser cosmic-ray model~\cite{Gaisser:2011cc} and the NRLMSISE-00 atmospheric model~\cite{Picone:2002go}. With this procedure, meson and proton fluxes are obtained as a function of $X$, $E$ and $\cos\theta$ (see  Appendix~\ref{app:fluxes} for details). At this point, it is worth mentioning that these fluxes are subject to significant theoretical uncertainties related to the choice of cosmic-ray and hadronic interaction models, see e.g. Fig.~9 in Ref.~\cite{Fedynitch:2012fs} or Ref.~\cite{Garzelli:2016xmx}, which can be of the order of 20-25\%. Additional uncertainties may arise from seasonal and geomagnetic effects. As discussed in Ref.~\cite{Honda:2015fha}, while seasonal effects are negligible at SK, they could be potentially large for the fluxes observed at Icecube. However, they tend to cancel out when you take data over a whole year (the impact of the seasonal effect for Icecube can be seen e.g. in Ref.~\cite{icecube-seasonal}). On the other hand, geomagnetic effects mostly affect the azimuthal distribution and horizontal events~\cite{Gaisser:2002jj}. Since we only study time-averaged zenith distributions, their effects in the azimuthal distributions are averaged out. For SK, the three-dimensional effects change the height distribution of atmospheric neutrinos predominantly in the horizontal direction rather than the vertical direction where our signal dominates; see Figs.~18 and 19 in Ref.~\cite{Honda:2004yz}. In the case of the IceCube analysis effects on the horizontal events are negligible as geomatic effects are only relevant for neutrino energies below ~20 GeV (see Fig.~11 in Ref.~\cite{Honda:2004yz}). 

\begin{figure}
\begin{center}
\includegraphics[width=0.8
\columnwidth]{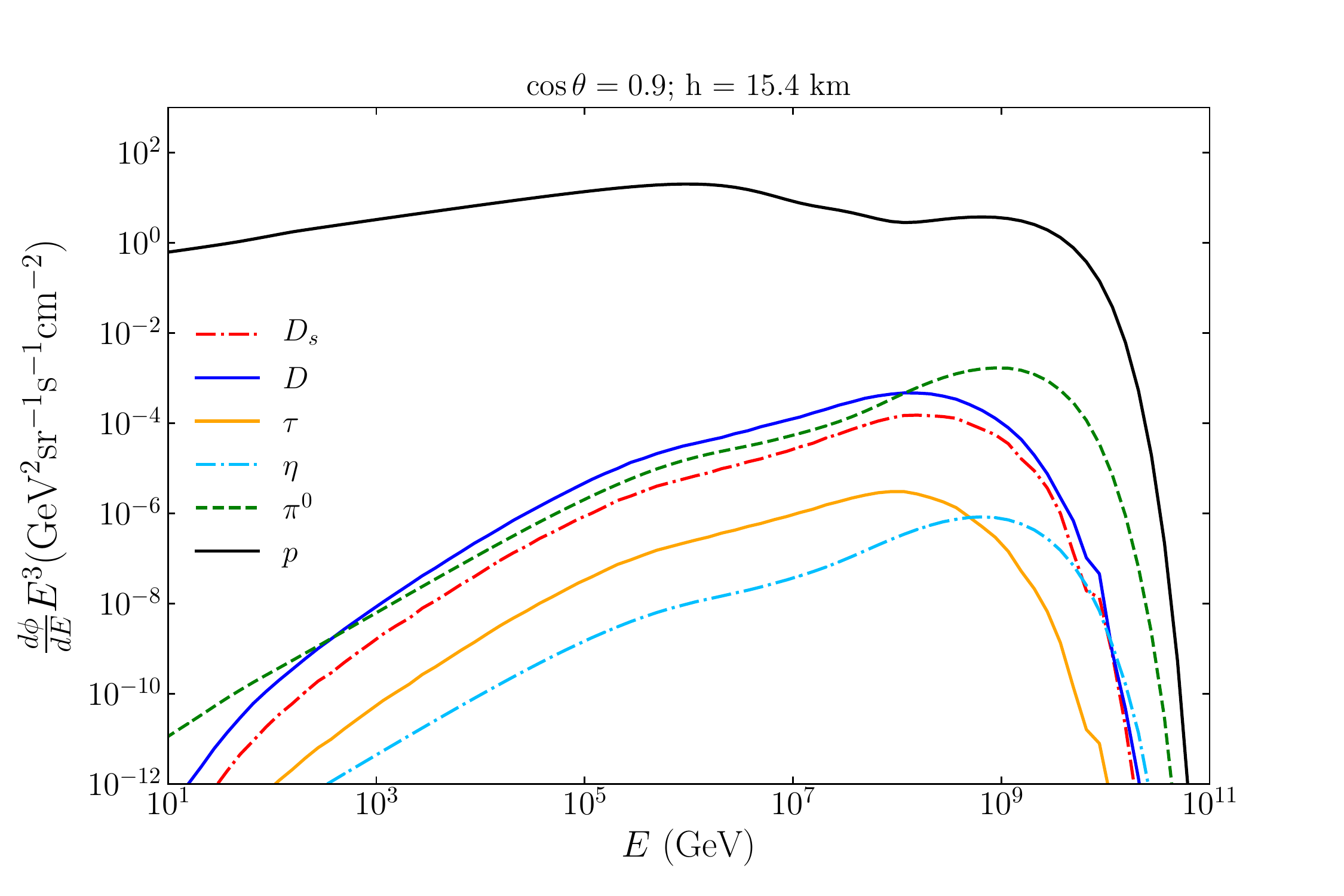} 
\end{center}
\caption{Differential fluxes of LLP parent particles as a function of the energy at fixed height in the atmosphere, $h=15.4$km. }
\label{fig:parentsE}
\end{figure}

\begin{figure}
\begin{center}
\includegraphics[width=0.8
\columnwidth]{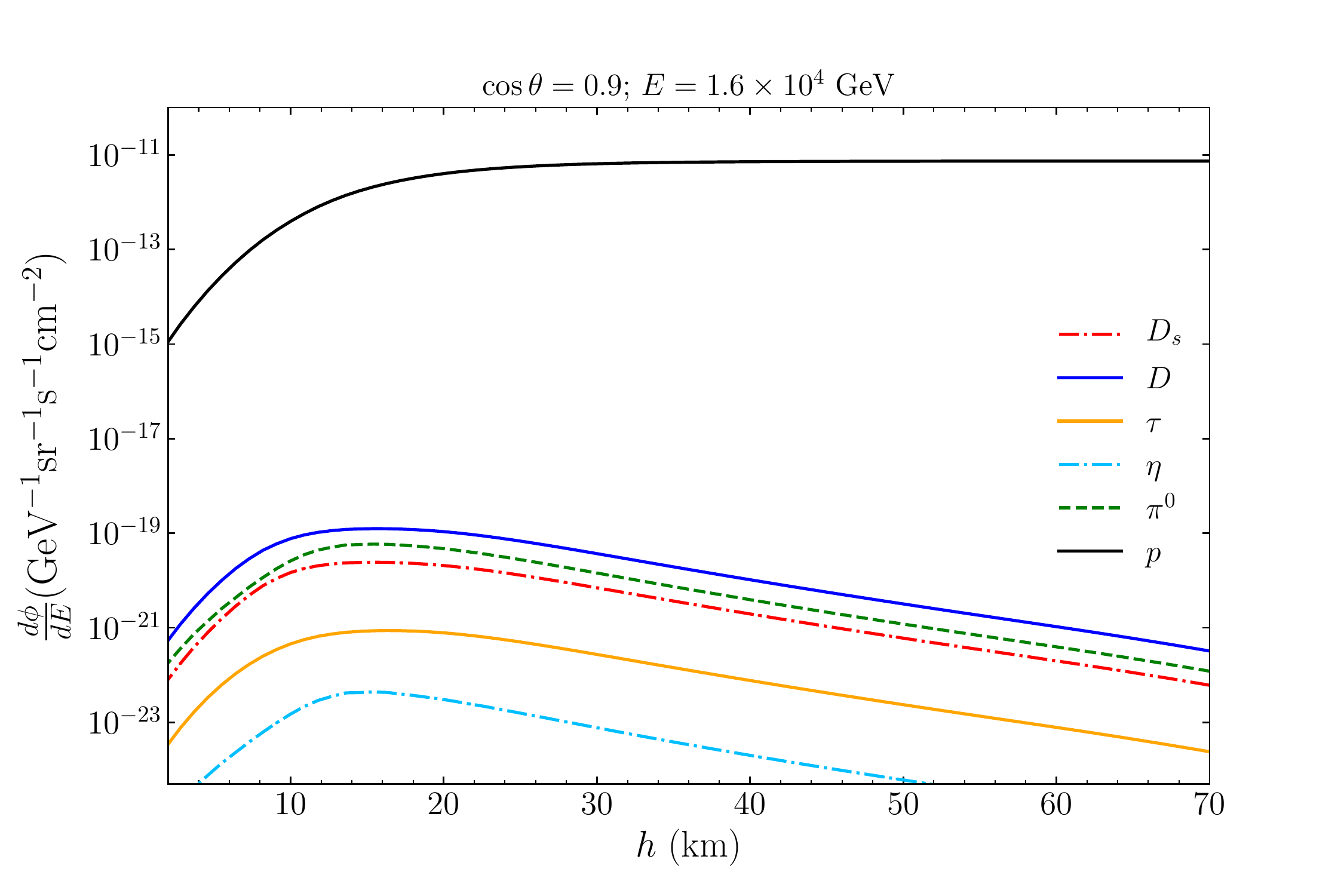} 
\end{center}
\caption{Differential fluxes of LLP parent particles as a function of the height at fixed energy, $E=1.56 \cdot 10^4$ GeV.}
\label{fig:parentsh}
\end{figure}

We will  assume that the boost is large enough $E \gg m_P$ so that the zenith angle of the LLP is that of the parent particle and the integration over  the parent zenith angle becomes trivial. In the following we will consider models where parent particles of LLP are protons and mesons. The differential fluxes of these particles as a function of the energy are shown in Fig.~\ref{fig:parentsE} at a fixed height, and as a function of the height at fixed energy in Fig.~\ref{fig:parentsh}.
%%%%%%%%%%%%%%%%%%%%%%%
%%%%%%%%%%%%%%%%%%%%%%%

%%%%%%%%%%%%%%%%%%%%%%%
%%%%%%%%%%%%%%%%%%%%%%%

In the rest of this section we discuss separately the different production mechanisms of LLPs,  from SM mesons and $\tau$ lepton decays, as well as from proton \bremm . In the case of production via meson decays, we will further distinguish between two- and three-body decays, as the outgoing LLP energy distributions will be different in the two cases.

\subsection{Production from decays of SM particles}

Let us first consider that a LLP, $A$, is produced in the decay of a SM meson $P$. Once the flux of the parent meson has been computed, the production profile of the LLP in the decay $P \rightarrow A Y $ or $P\rightarrow A Y Z$ is given by~\cite{Gondolo:1995fq}
\begin{eqnarray}
{d \Pi_A \over d E d \cos\theta d \ell} = \sum_{ch}  \int_{E_P^{\rm min}}^{E_P^{\rm max}} ~d E_P {1 \over  \gamma_P \beta_P c \tau_P} {d\Phi_P(E_P,\cos\theta)\over dE_P d\cos\theta }   {d n_{ch} (E_P, E)\over d E  }
\label{eq:master}
\end{eqnarray}
where the sum goes over all channels (labeled as $ch$) contributing to the LLP production from $P$ decays. Here, $d n_{ch}/dE$ is the fraction of decaying parents that produce an $A$ with energy in the bin $[E, E+dE]$ in the decay channel $ch$, while $\gamma_P$, $\beta_P$ and $\tau_P$ are the boost, velocity and lifetime in the rest frame of the parent particle $P$.

Both the distribution $d n_{ch}/dE$ and the integration limits for $E_P$ ($E_P^{\rm min}, E_P^{\rm max}$) depend on the decay kinematics and, in particular, they depend on whether the LLP is produced in a two- or in a three-body decay. 

\subsubsection{Two-body decays}

In two-body decays $P\rightarrow A Y$, the differential distribution $dn/dE$ is flat in $E$: 
\begin{eqnarray}
{d n (P\rightarrow A Y; E_P, E)\over d E  } \simeq {1\over \Gamma_P} {d\Gamma(P\rightarrow A Y) \over dE} =  {{\rm Br}(P\rightarrow A Y) \over \Gamma(P\rightarrow A Y)} {d\Gamma(P\rightarrow A Y) \over dE},
\label{eq:dn2b}
\end{eqnarray}
with 
\begin{eqnarray}
{1 \over \Gamma(P\rightarrow A Y)} {d\Gamma(P\rightarrow A Y) \over dE} = {1\over p_P \sqrt{\lambda(1,y_A^2, y_Y^2)}},
\label{eq:2bd}
\end{eqnarray}
where $y_i \equiv {m_i\over m_P}$ and 
\begin{eqnarray}
\lambda(a,b,c) \equiv a^2+ b^2 + c^2 -2 a b -2 a c -2 b c.
\label{eq:lambda}
\end{eqnarray}
The kinematical limits for the energy are:
\begin{eqnarray}
\gamma_P E_{\rm max} - {p_P \sqrt{\lambda(1,y_A^2, y_Y^2)}\over 2} \leq E \leq \gamma_P E_{\rm max} + {p_P \sqrt{\lambda(1,y_A^2, y_Y^2)}\over 2},
\end{eqnarray}
where $E_{\rm max}$ is the energy of the LLP $A$ when the decay takes place in the meson's rest frame, that is:
\begin{eqnarray}
E_{\rm max} \equiv {m_P\over 2} (1+ y_A^2-y_Y^2),
\end{eqnarray}
and $\gamma_P = E_P/m_P$. Alternatively if $E$ is fixed, the kinematical limits for $E_P \gg m_P$  are:
\begin{eqnarray}
{m_P E\over E_{\rm max} + {m_P \over 2} \sqrt{\lambda(1,y_A^2, y_Y^2)}} \leq E_P \leq {m_P E\over E_{\rm max} - {m_P \over 2} \sqrt{\lambda(1,y_A^2, y_Y^2)}}.
\end{eqnarray}

\subsubsection{Three-body decays}

In the case of three-body decay, $P \rightarrow A Y Z$, we have:
  \begin{eqnarray}
{\rm Br}(P \rightarrow A Y Z)^{-1} {d n (P\rightarrow A Y Z ; E_P, E)\over d E  }  =  {1 \over  \Gamma(P \rightarrow A Y Z)} {d\Gamma(P\rightarrow A Y Z) \over dE}.
\label{eq:dn3b}
\end{eqnarray}
The distribution is no longer flat and the kinematical limits, neglecting masses of $Y$ and $Z$, are:
\begin{eqnarray}
\gamma_P \left(E'_{\rm max} -  \sqrt{{E'}_{\rm max}^2 -m_A^2} \right)  \leq E \leq   \gamma_P \left(E'_{\rm max} +\sqrt{E_{\rm max}^{'2} -m_A^2} \right),
\end{eqnarray}
with
\begin{eqnarray}
\label{eq:Emax}
 E'_{\rm max} \equiv {m_A^2 + m_P^2 \over 2 m_P}.
\end{eqnarray}
Fixing $E$, the kinematical limits for $E_P$ are therefore:
\begin{eqnarray}
 {m_P E\over \left(E'_{\rm max} + \sqrt{{E'_{\rm max}}^2 -m_A^2} \right) } \leq E_P \leq   {m_P E\over \left(E'_{\rm max} -  \sqrt{{E'_{\rm max}}^2 -m_A^2} \right) }.
\end{eqnarray}

\subsection{\Bremm}

An alternative production mechanism for vectors and scalars is through \bremm, in the scattering of SM particles with air nuclei. 
Let us consider the case of a dark photon $V$ of mass $m_V$ emitted by a proton $p$. In analogy to Eq.~(\ref{eq:master}), the production profile of the dark photon can be written in this case as~\cite{Gondolo:1995fq}:
\begin{eqnarray}
{ d\Pi_V\over d E d\cos\theta d X} = \int_{E}^\infty d E_p {1 \over  \lambda_p(E_p)} {d \Phi_p(E_p,\cos\theta,X)\over d E_p d\cos\theta} {d n (p \mathcal{N} \rightarrow V Y; E_p, E)\over d E} ,
\label{eq:brem}
\end{eqnarray}
where $\Phi_p$ is the proton flux and  $\lambda_p$ is the interaction thickness of protons in air, \ie the average amount of atmosphere (in g/cm$^2$) traversed between successive collisions with air nuclei $\mathcal{N} $.

It is common to parametrize the coupling of the dark photon to the proton as the QED coupling multiplied by a small number $\epsilon$. In this case, the differential distribution $dn/dE$ takes the form \cite{Blumlein:2013cua,Gorbunov:2014wqa,deNiverville:2016rqh}:
\begin{eqnarray}
 {d n (p \mathcal{N} \rightarrow V Y; E_p, E)\over d E} = {1 \over E_p} \int_0^{p^{2 max}_{\perp}} d p_\perp^2 {\sigma_{p \mathcal{N} }(2 m_p (E_p-E)) \over \sigma_{p \mathcal{N} }(2 m_p E)} w(z, p_\perp^2),
 \label{eq:bremdn}
\end{eqnarray}
where  $|p_{\perp}^{\rm max}|  = 1$GeV and 
\begin{eqnarray}
w(z, p^2_\perp) \equiv {\epsilon^2 \alpha_{QED} \over 2 \pi H} \left[ {1 + (1-z)^2\over z} - 2 z (1-z) \left({2 m_p^2 + m_V^2 \over H} - z^2 {2 m_p^4 \over H^2}\right)\right.\nonumber\\
+ \left. 2 z (1-z) (z + (1-z)^2) {m_p^2 m_V^2 \over H^2} + 2 z (1-z)^2 {m_V^4 \over H^2} \right] \, .
\end{eqnarray}
Here, $m_p$ is the mass of the proton and 
\begin{eqnarray}
H(z, p_\perp^2) \equiv p_\perp^2 + (1-z) m_V^2 + z^2 m_p^2,
\end{eqnarray}
with $z \equiv p_V/p_p$, where $p_V$ and $p_p$ are the dark photon and proton momenta, respectively.

\section{Expected number of LLP decays in atmospheric neutrino detectors}
\label{sec:detect}

Once the production profile of the LLP has been computed, the flux of particles that arrive at the detector $\Phi_A$ is simply obtained integrating over all LLPs produced at different distances $\ell$ from the detector, weighted by their corresponding survival probabilities, as 
\begin{eqnarray}
{d \Phi_A\over dE_A d\cos\theta} = \int_0^{\ell_{\rm max}} d \ell {d \Pi_{A}\over dE_A d \cos\theta d\ell} 
~e^{-{\ell\over \ell_{\rm decay}}},
\label{eq:fluxA}
\end{eqnarray}
where $d \Pi_A$ is the flux of LLP produced at a distance $\ell$ of the detector within the interval $[\ell,\ell+d\ell]$, computed as outlined in Sec.~\ref{sec:production}, and $\ell_{\rm decay}$ is the decay length of the LLP in the lab frame, related to its lifetime ($\tau_A$) as
\begin{eqnarray}
\ell_{\rm decay} = \gamma_A \beta_A c \tau_A \simeq {E_A\over m_A} c \tau_A ,
\end{eqnarray}
where $\beta_A, \gamma_A$ are the boost parameters of the LLP, and we assume $E_A \gg m_A$. 
We will assume that, once the particle is produced, since it is very weakly coupled it does not interact any further in the atmosphere and only decays.
Also, note that the upper limit of the integral in Eq.~\eqref{eq:fluxA} is in practice a function of $\theta$ (see Eq.~\eqref{eq:ell}): $ \ell_{\rm max} \equiv \ell(h_{\rm max},\theta)$ and $h_{\rm max} \simeq 80$km is the maximum height of the atmosphere where cosmic showers are produced.

\begin{figure}[ht!]
\begin{center}
\includegraphics[width=0.7\columnwidth]{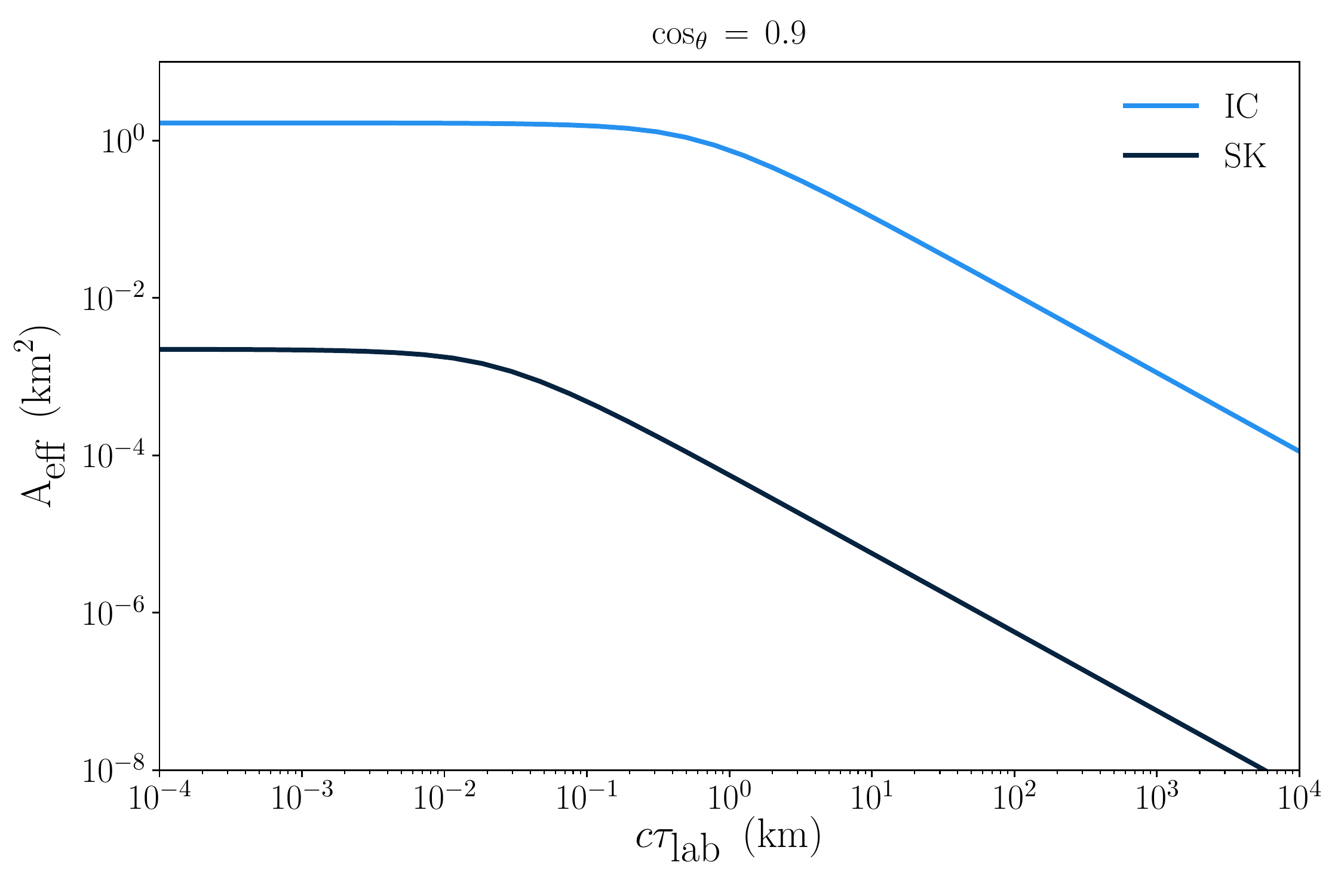} 
\end{center}
\caption{Effective decay area for the IceCube detector ($R_{IC}={1 \over \sqrt{\pi}}$ km and $H_{IC} = 1$km) and Super-Kamiokande ($H_{SK}=0.04$~km and $R_{SK}=0.02$~km) as a function of the decay length in the laboratory frame for down-going events ($\cos\theta=0.9$).}
\label{fig:aeff}
\end{figure}

Using the flux in Eq.~\eqref{eq:fluxA}, the number of decays inside the detector within a given time window $\Delta T$, for LLPs with energies and trajectories in the intervals $[E_A, E_A + dE_A]$, $[\cos\theta, \cos\theta + d\cos\theta]$, can be computed as
\begin{equation}
\frac{d N}{dE_A d\cos\theta} = \Delta T  A^{\rm eff}_{\rm decay}(E_A,\cos\theta) {d \Phi_A\over dE_A d\cos\theta },
\end{equation}
where $ A^{\rm eff}$ is an effective area which accounts for the probability that a decay takes place inside the detector. This area can be estimated integrating the surface of the detector normal to the flux direction, weighted by the decay probability of the LLP inside the detector:
\begin{eqnarray}
A^{\rm eff}_{\rm decay}(E_A,\cos\theta) = \int dS_{\perp} \left\{ 1- \exp \left(- { \Delta\ell_{det} (\cos\theta)\over \ell_{\rm decay}(E_A) } \right) \right\} \, .
\end{eqnarray}
Here, $\Delta\ell_{det}$ is the length of the segment of the LLP trajectory that cuts into the detector. Its calculation, which just depends on the zenith angle defining the trajectory of the LLP and on the geometry of the detector, is outlined in App.~\ref{app:decay}. Figure~\ref{fig:aeff} shows the effective decay areas for the IceCube and Super-Kamiokande detectors as a function of the lifetime of the LLP in the lab frame. 

%%%%%%%%%%%%%%%%%%%%%
%%%%%%%%%%%%%%%%%%%%%

\subsection{Detector efficiencies and datasets used}
%\label{sec:detector}

The number of events will also depend on the detector efficiencies and reconstruction effects. In this work, two atmospheric neutrino detectors have been considered: IceCube (IC) \cite{Abbasi:2008aa} and Super-Kamiokande (SK) \cite{Fukuda:2002uc}. Since they observe events in two very different energy regimes, we expect their results to be complementary and to probe different regions in parameter space. In this section we describe the assumptions and data sets used for each of the two detectors separately.

\subsubsection{IceCube}

The IceCube Neutrino Observatory~\cite{Abbasi:2008aa} is a $\sim 1$~km$^3$ neutrino detector at the Geographic South Pole, optimized for detecting neutrinos with energies above 100~GeV. For the effective area, Eq.~(\ref{eq:aeff}), we consider a simplified cylindrical geometry with $H=1$~km and volume 1~km$^3$. 
The effective area as a function of the  decay length in the laboratory frame is shown in Fig.~\ref{fig:aeff}.  

We consider the data sample corresponding to the analysis presented in Ref.~\cite{Aartsen:2014muf}, for which effective areas for  interacting neutrino events are publicly available in the IceCube collaboration webpage~\cite{icecubeweb}. The statistics corresponds to 641 days of contained events, and the reconstructed data samples are divided into tracks and cascade events, in the northern and southern hemispheres, and are binned in reconstructed energy. Effective areas for each of these samples are provided for each particle, interaction type and true neutrino energy. From these effective areas we can infer the detector efficiencies by dividing by the corresponding interaction cross sections. We therefore estimate the detector efficiencies
as
\begin{eqnarray}
\epsilon^{\alpha\beta}_\nu(E^\nu_{ \rm rec}, E^\nu_{ \rm true},\theta_{\rm true}) \simeq {A_{{\rm eff} IC}^{\alpha\beta}(E^\nu_{ \rm rec}, E^ \nu_{ \rm true},\theta_{\rm true})\over \rho_{\rm ice} V_{IC} N_A \sigma^\nu_\beta(E^\nu_{\rm true})}
\end{eqnarray}
where $\alpha$ selects the reconstructed sample: cascade/track and north/south, $\beta$ selects the true neutrino interaction type, CC-e, NC, etc. In the denominator we have the total cross section in the IceCube detector $\sigma^\nu_\beta$, which is estimated in terms of the nucleon-neutrino cross section, times the number of nucleons in the detector: Avogadro number $N_A$, times the ice density ($\rho_{\rm ice}\simeq 0.92~{\rm gr~cm^{-3}}$) times the fidutial volume $V_{IC}$. The required neutrino-nucleon cross sections are extracted from Ref.~\cite{CooperSarkar:2011pa}. 

We will make the assumption that our LLP decay events have the same efficiencies as charged-current (CC) neutrino interactions with a similar set of observed particles in the final state. We also assume that the relationship between the energy of the decaying particle and the deposited energy of all particles produced in the detector is similar to the relationship between true and reconstructed neutrino energies in CC interactions. We believe that this is a reasonable as long as all particles produced in the LLP decay are visible in the detector. More concretely we will consider only LLP visible decays to electrons and charged hadrons, mainly because the expected neutrino background is smaller, they are more likely to be contained, and their the energy can be better determined. We assume that LLP decays of this type have an efficiency similar to those of CC e-like neutrino interactions:
\begin{eqnarray}
\epsilon^{\alpha ~{\rm e-like, hadronic}}_{\rm LLP}(E_{\rm rec}, \theta_{\rm rec}; E_A,\theta) &\simeq & \epsilon_\nu^{\alpha ~{\rm CC-e}}(E_{\rm rec},\theta_{\rm rec}; E_A,\theta).
\end{eqnarray}
In our IC calculations, we use the efficiencies corresponding to $\alpha=$cascades. Therefore,  we weight the number of signal events by the branching ratio into \emph{e-like events}, in order to consider all decay channels which have electrons, photons, $\tau$-leptons or hadronic showers in the final state, which would be observed as cascade-like events. 

An important point is that these effective areas {\it do not} include the muon veto applied in the event pre-selection. This veto reduces events with a muon signal within a $3\mu s$ time window of the neutrino signal.  
This veto is designed to optimize the search for astrophysical neutrinos, as it reduces significantly the overwhelming background from atmospheric muons, and the atmospheric neutrino background in southern sky events \cite{Schonert:2008is}. At sufficiently high energies it is quite likely that muons from the same atmospheric shower that produce the LLPs will also reach the detector within this large time window, and therefore it also reduces significantly our signal. Since this veto has been applied to the background and the data, we need to impose it also on our signal to extract bounds, however it is quite likely that an optimized LLP search can be carried out, where the time window for a coincident muon track is reduced significantly, since the LLP has an atmospheric origin and an accompanying muon is not unexpected. The veto is expected to affect similarly our signal and the prompt neutrino signal, which is the dominant background, so it does not improve the signal-to-noise ratio of the LLP search. To implement this veto on our signal we use the passing fractions of the muon veto, ${\mathcal P}_{\rm pass}(E_A,\theta)$ obtained in Ref.~\cite{Arguelles:2018awr}, which depend on the true energy and zenith of the LLP. Given that we are concentrating on  cascade events, and in most LLP production there are no associated muons, the most appropriate passing fractions  to use are those corresponding prompt $\nu_e$. 

The number of signal events in the $i$-th energy bin at IC can be finally obtained as
\begin{eqnarray}
N^{\rm IC}_i = {\rm Br}(A\rightarrow e{\rm-like}) \! \int  \! d\cos\theta \! \int dE_A  ~
\epsilon_\nu^{\alpha \beta}(E_i, \theta_{\rm rec}; E_A, \theta) {\mathcal P}^{\rm prompt- \nu_e}_{\rm pass}(E_A,\theta)~ \frac{dN}{dE_A d\cos\theta} \, , \nonumber \\
\label{eq:nevents}
\end{eqnarray}
where $0.2 \leq \cos \theta_{\rm rec}\leq 1$ corresponds to the southern sky, while $\alpha = {\rm cascade}$ and $\beta=$CC $e$-like.

It might be useful to also include DeepCore atmospheric data. However, the detector volume in this case is much smaller than for IceCube, and the energy range considered for atmospheric neutrino analyses for DeepCore is comparable to that of the multi-GeV data in SK. Therefore, we expect similar results for DeepCore as those obtained for SK. It may however be interesting to search for a signal using atmospheric neutrinos with energies between 100~GeV and 1~TeV, which may be sensitive to a different range of values of $c\tau$. This is unfortunately not possible using currently available public data.

\subsubsection{Super-Kamiokande}

Super-Kamiokande is a significantly smaller water Cherenkov detector, but it has a lower energy threshold where the atmospheric flux is much higher and therefore might have a comparable or event better sensitivity than IceCube to the scenarios considered. In the case of SK, in the computation of the effective decay areas (Eq.~(\ref{eq:aeff})) we have assumed a cylindrical geometry with $H=0.04$~km and $R=0.02$~km. The result is shown in Fig.~\ref{fig:aeff}.

We will again only consider electron-like events, that is, events with an e-like topology as defined above. We have used in this case the fully-contained multi GeV e-like sample from Ref.~\cite{Abe:2017aap} (including single and multi-ring) with reconstructed zenith angle $0 \leq \cos\theta \leq 1$. We extract data as well as neutrino background  from Fig.~5 of \cite{Abe:2017aap}. In this case, while we do have information regarding the zenith angle of the event sample (see Fig.~5 in Ref.~\cite{Abe:2017aap}) we do not have available information regarding the energy of the events. Therefore, for SK we bin the data only in $\cos\theta$ and integrate over all neutrino energies between 1~GeV and 90~GeV, which is the range of parent neutrino energies for multi-GeV e-like contained events (see Fig.~6 of  Ref.~\cite{Abe:2017aap}). We think this is conservative, as SK may be sensitive to events outside this range. We also assume that the efficiencies for the decay of the LLP are similar to those of electron neutrino CC events in the multi-GeV range. Our only assumption regarding detection efficiencies for Super-Kamiokande is that they are flat as a function of energy, which we think is a reasonable assumption for fully contained events. We note that migration matrices for SK are not publicly available. 

The number of events in the $i$-th bin in $\cos\theta$ can therefore be computed as:
\begin{eqnarray}
N^{\rm SK}_i = {\rm Br}(A, e{\rm-like}) \! \int_{\cos\theta_{i}^{\rm min}}^{\cos\theta_i^{\rm max}} \! d\cos\theta \! 
\int_{1~\rm{GeV}}^{90~\rm{GeV}} dE_A  
\epsilon^{SK} \frac{dN}{dE_A d\cos\theta} \, , \nonumber \\
\label{eq:nevents}
\end{eqnarray}
where $\cos\theta_{i}^{\rm min}$ and $\cos\theta_{i}^{\rm max}$ are the lower and upper limits of the bin. In this case, we have assumed a flat detection efficiency $\epsilon^{SK} = 0.75$, in line with the values quoted in Ref.~\cite{Abe:2017aap} for the multi-GeV $\nu_e$ event sample.

\section{Atmospheric LLP in selected scenarios}
\label{sec:LLP}

We now consider three simple BSM scenarios of very weakly interacting sectors that can lead to LLPs. In the first two examples, both production and decay are controlled by the same couplings, in such a way that the requirements of long enough lifetimes and large enough production go in opposite directions. Instead in the third example we consider a scenario where it is possible to decouple production from decay.

\subsection{Heavy Neutral Leptons}
\label{sec:HNL}

The existence of heavy neutral leptons (HNL) is a  generic prediction of  Type-I seesaw models of neutrino masses \cite{Minkowski:1977sc,GellMann:1980vs,Yanagida:1979as,Mohapatra:1979ia}. The model is the simple extension of the SM with at least two heavy Majorana singlets $N_j$. In the basis where the Majorana mass terms are diagonal, the Lagrangian reads:
\begin{eqnarray}
 {\mathcal L}_N =  {\mathcal L}_{SM} +  \sum_j i \bar{N}_j \gamma^\mu  \partial_\mu N_j - \left(Y_{\alpha j} \bar{L}_\alpha {\tilde\Phi} N_j + {m_{N_j} \over 2} \bar{N}_j N_j^c \right) \, ,
\label{eq:VNmodel}
\end{eqnarray}
where $\tilde\Phi \equiv i\sigma_2 \Phi^*$ is the complex conjugate of the Higgs field $\Phi$, $L_\alpha$ is the SM lepton doublet with flavor $\alpha$, $Y_j$ is a Yukawa coupling and $m_{N_j}$ is the Majorana mass of the singlet $N_j$, which in principle is a free parameter of the model.   
After spontaneous symmetry breaking, the heavy Majorana states mix with the standard neutrinos resulting in a spectrum of almost standard light states (with masses $m_\nu \propto {(Y v)^2\over m_N}$) which correspond to the light neutrinos, and almost singlet 
heavy states corresponding to the HNL (with masses $\propto m_N$). In the following, the full leptonic mixing matrix will be denoted as $U$ and the mixing between the charged leptons and the heavy states is given by $|U_{\alpha j}|^2$, which naively scales like $\left({Y v \over m_N}\right)^2$. 
It is through this mixing that the heavy singlets could be produced either through CC or neutral-current (NC) processes and also how they would decay back to SM particles. 

A priori, the Majorana masses of the HNL are free parameters of the theory and therefore have to be probed experimentally. The GeV range is interesting from the theoretical point of view, since it has been shown that models with HNL at the GeV scale could explain neutrino masses and mixing parameters, as well as the matter-antimatter asymmetry of the Universe~\cite{Akhmedov:1998qx,Asaka:2005pn} without conflicting cosmological bounds. The search for GeV scale HNL is a very active field and several future experiments such as SHIP \cite{SHiP:2018xqw}, FCCee \cite{Blondel:2014bra}, or DUNE \cite{Ballett:2019bgd} have the potential to improve significantly over present bounds \cite{Atre:2009rg, Drewes:2015iva}. On the phenomenological front, new avenues to constrain these models are being proposed and/or further explored using current and past data~\cite{Dib:2019tuj,Kobach:2014hea,Bryman:2019bjg,Bryman:2019ssi,Coloma:2017ppo,Coloma:2019qqj,Chun:2019nwi}. 

Here we consider instead the production of HNL in atmospheric showers, where their dominant production mechanism is through meson decays. Depending on their mass, the leading production channel is $\pi$, $K$, $D_{(s)}$-meson decays (or in the decays of even heavier resonances, such as $B$ mesons). Particularly interesting is the mass range slightly above the kaon mass, where existing laboratory bounds are weaker~\cite{Atre:2009rg,Drewes:2015iva,Bryman:2019bjg,Bryman:2019ssi}. We have therefore considered the production via $D$- and $D_s$-meson decays, as well as from $\tau$ decays if one considers the possibility that the production is dominated by the poorly constrained coupling to taus. The production in the lower-mass range from kaon and pion decays will be considered elsewhere. For simplicity, hereafter we use the notation $U_{\alpha} \equiv U_{\alpha j}$ ($\alpha = e,\mu,\tau$) to refer to the elements of the leptonic mixing matrix that control both the production and decay of the heavy state $N$.

\begin{figure}
\begin{center}
\includegraphics[width=0.8\columnwidth]{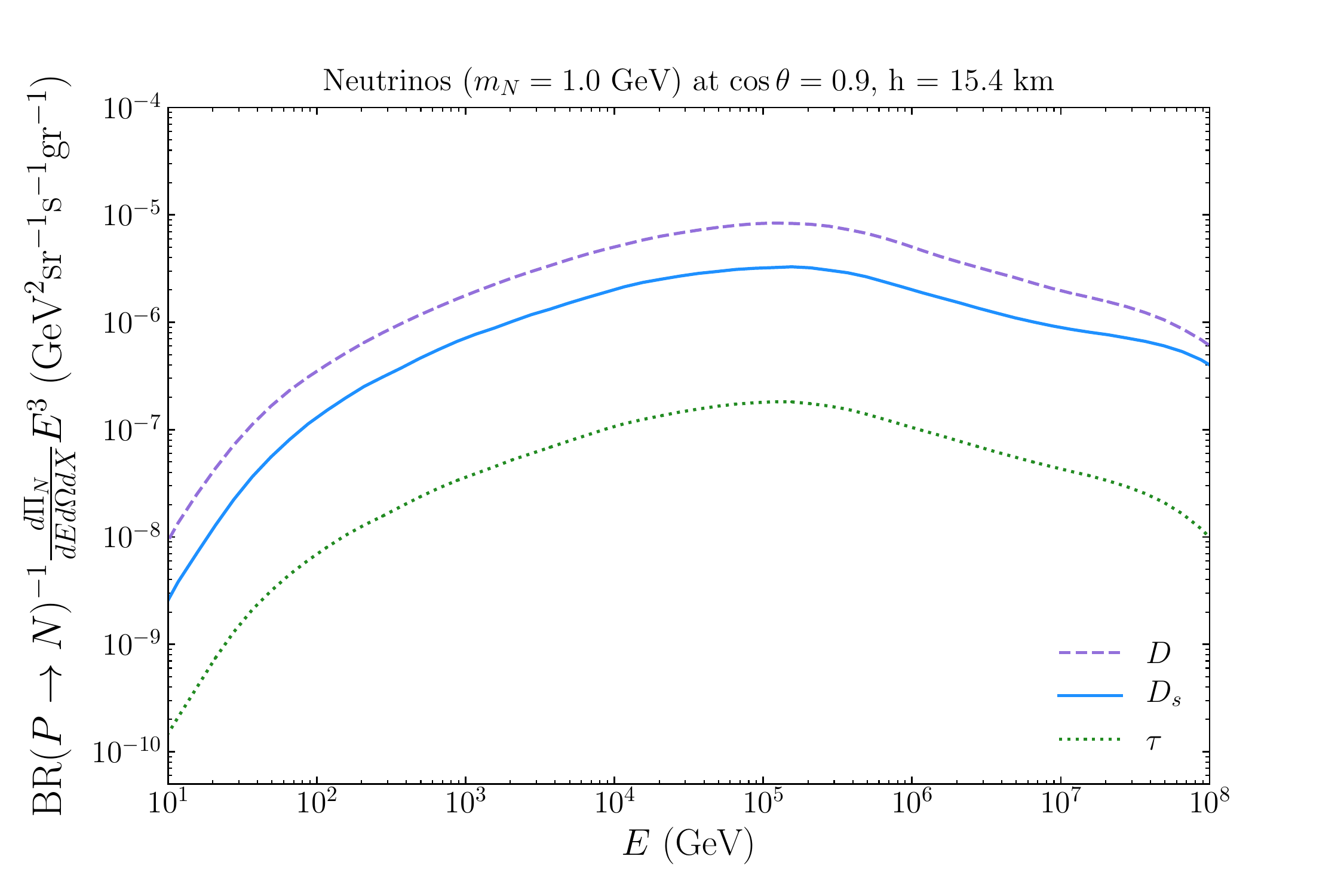}  
\end{center}
\caption{Production profile of HNL in the atmosphere, shown for $D, D_{s}$ and  $\tau$ decays separately, in units of the corresponding branching ratios and for $m_N=1$GeV. The profile is shown at approximately 15~km from the Earth's surface, since this is where the parent meson fluxes are expected to be maximal, see Fig.~\ref{fig:parentsh}, and for $\cos\theta = 0.9$ (we expect the signal to be largest as $\cos\theta \to 1$).
}
\label{fig:fluxesHNL}
\end{figure}

For $D_{(s)}$ meson decays, and for HNL between 0.5 and 1.5~GeV the dominant decay is two-body: $D^\pm \rightarrow N l^\pm_\alpha$, where the flavour of the lepton $\alpha= e, \mu$ (note that, if $m_N \geq m_K$, this decay with $\alpha=\tau$ is kinematically forbidden). The $N$ fluxes are obtained from Eqs.~(\ref{eq:master}), (\ref{eq:dn2b}) and (\ref{eq:2bd}). The production profile of $N$ from $D$ and
$D_s$ decays are shown in Fig.~\ref{fig:fluxesHNL} in units of the corresponding branching ratios into this channel.

%%%%%%%%%%%%%%%%%%%%%%%%%%
%%%%%%%%%%%%%%%%%%%%%%%%%%

In the case of taus, the decays to produce $N$ are necessarily three-body. Assuming that the dominant mixing is to $\tau$ leptons and the mass of the HNL is in the GeV range, the dominant decay channels are
 $\tau^\pm \rightarrow N \nu_\alpha l^\pm_\alpha$, where $l_\alpha \equiv e,\mu$. 
 In this case it is more convenient to perform the change of variables $E \to z\equiv E/\gamma_\tau$, where $\gamma_\tau = E_\tau/m_\tau$ is the boost factor of the $\tau$. Equation~(\ref{eq:dn3b}) then becomes 
  \begin{eqnarray}
{\rm Br}(\tau \rightarrow N \nu_\alpha l_\alpha) ^{-1} {d n (\tau\rightarrow N \nu_\alpha l_\alpha; E, E_\tau)\over d E  } =  {1\over \gamma_\tau \Gamma(\tau \rightarrow N \nu_\alpha l_ \alpha)} {d\Gamma(\tau\rightarrow N \nu_\alpha l_\alpha) \over dz} \, .
\end{eqnarray} 
The differential decay width in the rest frame of the $\tau$ lepton can be found in Ref.~\cite{Shrock:1981wq}, or in Appendix B of Ref.~\cite{Gorbunov:2007ak}. Neglecting all lepton masses except those of the $\tau$ and $N$, after boosting to the laboratory frame we get \footnote{Here we assume $\gamma_\tau m_N \geq E_{\rm max}$, where $E_{\rm max}$ can be found in Eq.~\eqref{eq:Emax}.} 
\begin{eqnarray}
{1\over \Gamma(\tau \rightarrow N \nu_\alpha l_ \alpha)} {d \Gamma(\tau \rightarrow N \nu_\alpha l_ \alpha)\over d z} = {1\over 72  z^3 X_\tau} &  (m_\tau-z)& \Big[m_N^6 (4 m_\tau^2-5 m_\tau z-5 z^2)  \nonumber\\
 & & - 9 m_N^4 m_\tau^2 z (m_\tau-3 z)  \nonumber\\
 & & + 9 m_N^2 m_\tau^2 z^3 (z-3 m_\tau) \nonumber \\
& & +m_\tau^3 z^3 (5 m_\tau^2+5 m_\tau z-4 z^2)\Big]  \nonumber \\
& & + {\mathcal O}(1-\beta_\tau) \, , 
\end{eqnarray}
where  
\begin{eqnarray}
X_\tau \equiv {-m_N^8+8 m_N^6 m_\tau^2-24 m_N^4 m_\tau^4 \log({m_N\over m_\tau})-8 m_N^2 m_\tau^6+m_\tau^8\over 24 m_\tau} .
\end{eqnarray}
The contribution to the flux from $\tau$ decays is also shown in Fig.~\ref{fig:fluxesHNL} in units of the  $\tau \rightarrow N l_\alpha \nu_\alpha$ branching ratio.  

The partial decay widths of $D$, $D_s$ and $\tau$ to $N$ can be obtained  from Refs. ~\cite{Shrock:1981wq}, ~\cite{Gorbunov:2007ak}, and~\cite{Shrock:1980ct}. For the $D$ and $D_s$ mesons, the total branching ratio for two-body decays with a HNL in the final state is obtained adding the contributions from $D_{(s)} \rightarrow N e$ and $D_{(s)} \rightarrow N \mu$. Assuming that the partial decay width of the parent meson to $N$ is very small compared to its SM decay width, the branching ratio reads
\begin{equation}
\label{eq:D-BR}
{\rm Br}(D_{(s)}\rightarrow N ) \simeq {G_F^2 f_{D_{(s)}}^2  |V_{D_{(s)}}|^2 m_{D_{(s)}}^3\over 8 \pi \Gamma_{D_{(s)}}} 
\sum_{\alpha = e,\mu} |U_{\alpha}|^2
\left[y_N^2+ y_{l_\alpha}^2- (y_N^2 -y_{l_\alpha}^2)^2  \right] \sqrt{\lambda(1,y_N^2, y_{l_\alpha}^2)},
\end{equation}
where $\Gamma_{D_{(s)}}$ is the total decay width of the $D_{(s)}$ meson (taken from Ref.~\cite{Tanabashi:2018oca}), $y_i \equiv m_i / m_{D_{(s)}}$ and $\lambda$ is defined in Eq.~(\ref{eq:lambda}). In Eq.~\eqref{eq:D-BR} $G_F$ is the Fermi constant, $f_{D_{(s)}}$ is the decay constant of the parent meson and $ V_{D_{(s)}}$ is the mixing matrix element in the CKM matrix that participates in the decay vertex. The values used for the meson masses and decay constants ($ f_D = 212$~MeV; $f_{D_s} = 249$~MeV) have been taken from Ref.~\cite{Tanabashi:2018oca}. For the CKM matrix elements, we use $V_D \equiv V_{cd} = 0.22$ and $V_{D_{s}} \equiv V_{cs}= 0.995$.
In the case of decays from taus, we assume that the $\tau$ branching ratio into $N$ is dominated by the value of $U_\tau$ since it is less constrained by other experiments. In this case, we get:
\begin{eqnarray}
{\rm Br}(\tau^\pm \rightarrow N) & = &\sum_{\alpha = e,\mu}{\rm Br}(\tau^\pm \rightarrow N l_\alpha^\pm \nu_\alpha) = 
\nonumber \\
& = & 2 {|U_\tau|^2\over 192 \pi^3} {G_F^2 m_\tau^5\over \Gamma_\tau} \left[1 - 8 y_N^2 - 24 y_N^4 \log(y_N)+8 y_N^6- y_N^8\right],
\end{eqnarray}
where, again in this case, for the total value of the width of the $\tau$ we use the SM value of $\Gamma_\tau$ (\ie  ignoring the small contribution of the HNL). 

The expected number of decays inside the volume of the detector will also depend on the lifetime
of the $N$, as outlined in Sec.~\ref{sec:detect}. The total width is computed adding the partial widths into two-body decay channels (that is, into charged mesons and charged leptons, or neutral mesons and neutrinos), the three-body decay channels into charged leptons and neutrinos, and the three-body invisible decay into three light neutrinos.
The partial widths for these decays can be found in several references \cite{Gorbunov:2007ak,Atre:2009rg,Bondarenko:2018ptm,Ballett:2019bgd} although they do not all agree. We have re-computed the two-body decay widths into mesons and leptons in the effective Lagrangian at low energies, and found good agreement with the results from the recent review in Ref.~\cite{Bondarenko:2018ptm}. 
The total width can be written as
\begin{eqnarray}
\Gamma_{\rm tot} =\sum_{\alpha=e,\mu,\tau} \Gamma_\alpha,
\end{eqnarray}
where the dependence on the mixing matrix elements goes as $\Gamma_\alpha \propto |U_{\alpha}|^2$. 
%%%%%%%%%%%%%%%%%%%
\begin{figure}[ht!]
\begin{center}
\includegraphics[width=10cm]{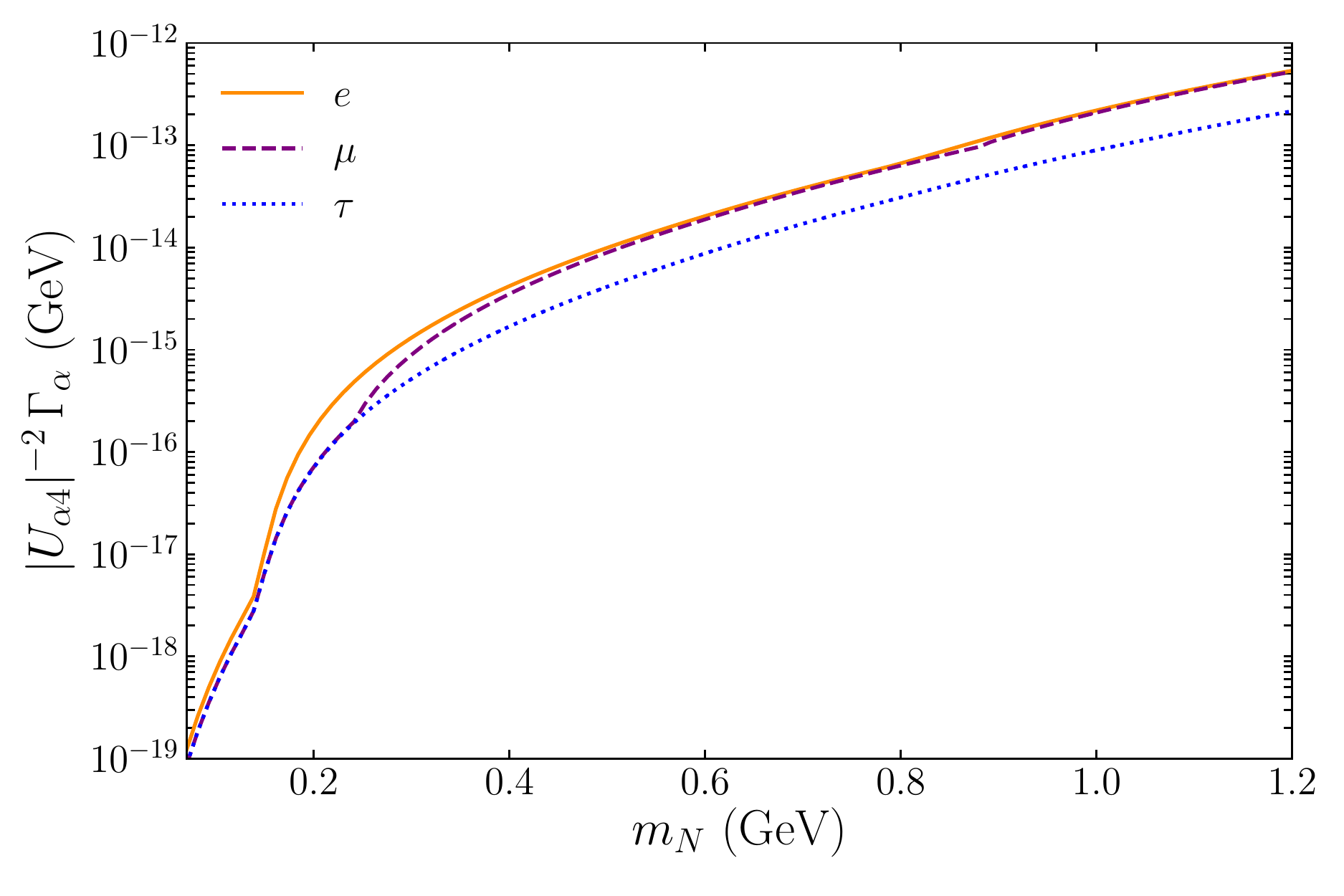} 
\end{center}
\caption{Ratios $|U_{\alpha}|^{-2} \Gamma_\alpha$ for $\alpha=e,\mu,\tau$ as a function of $m_N$. }
\label{fig:galpha}
\end{figure}
%%%%%%%%%%%%%%%%%%%
In Fig.~\ref{fig:galpha} we show the ratios $|U_{\alpha}|^{-2} \Gamma_\alpha$ as a function of the $N$ mass. From these, it is easy to reconstruct the total width and lifetime for any given values of the mixing matrix elements $|U_{\alpha}|^2$. 

\begin{figure}
\begin{center}
\includegraphics[width=4.9cm,keepaspectratio]{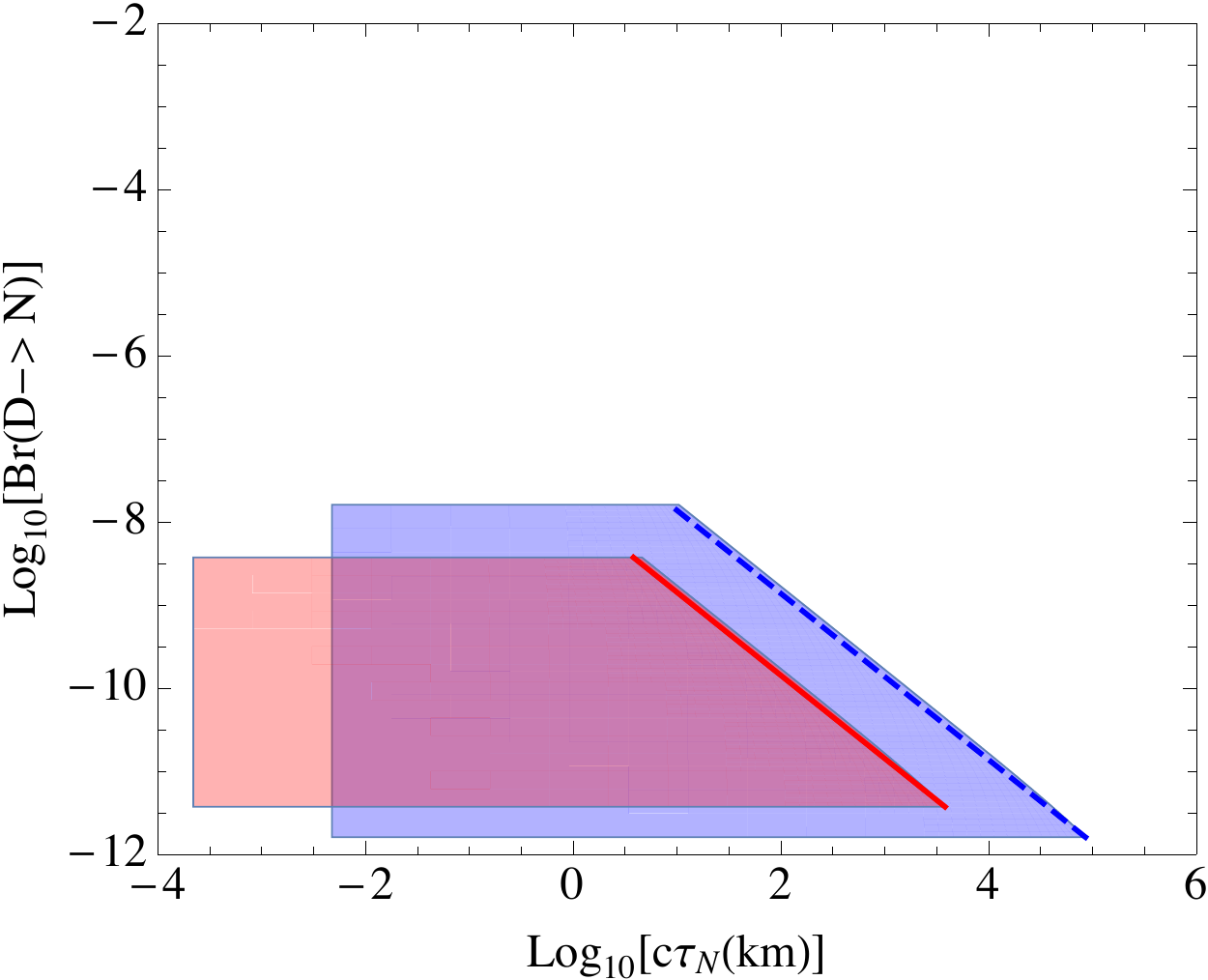}  
\includegraphics[width=4.9cm,keepaspectratio]{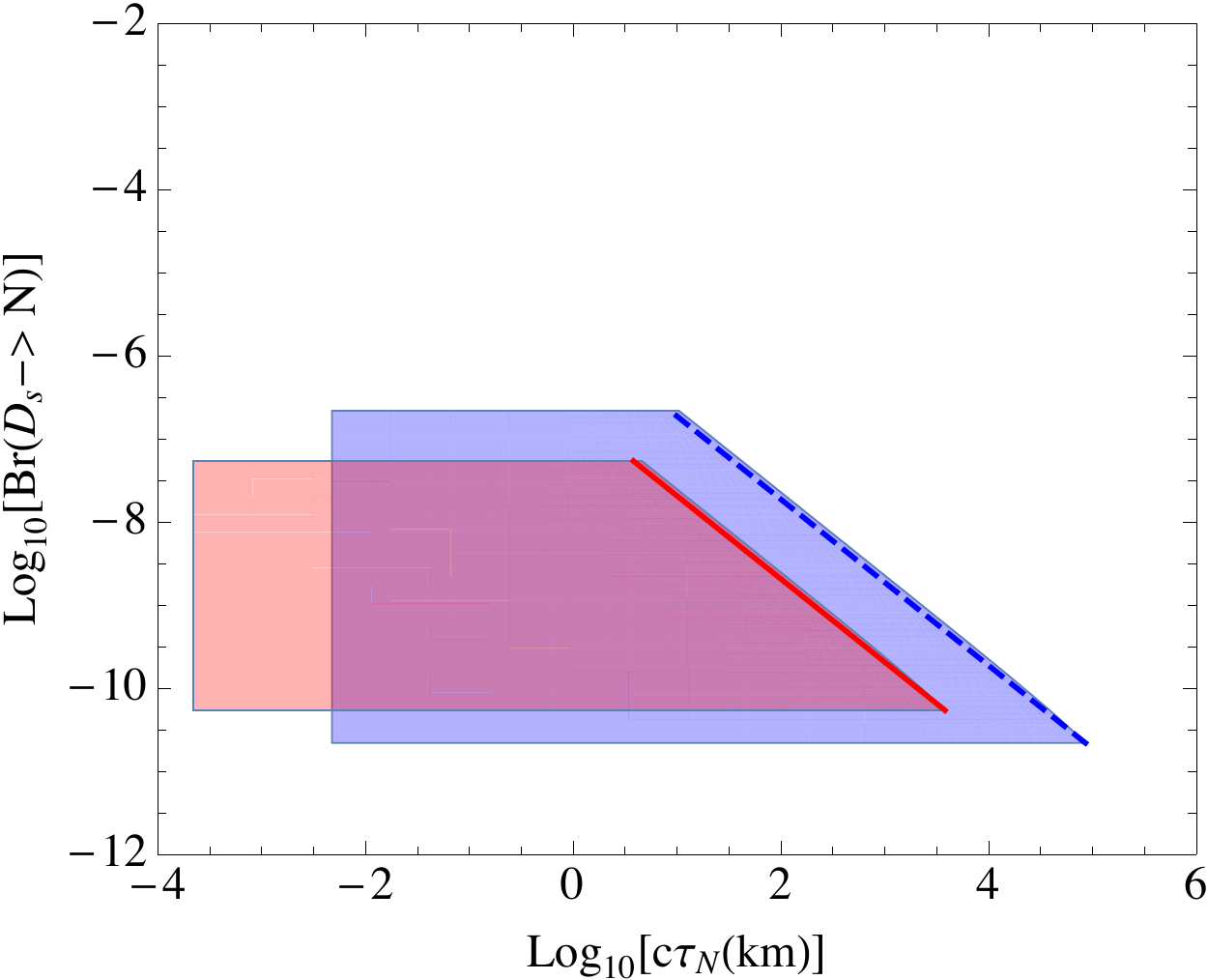} 
\includegraphics[width=4.9cm,keepaspectratio]{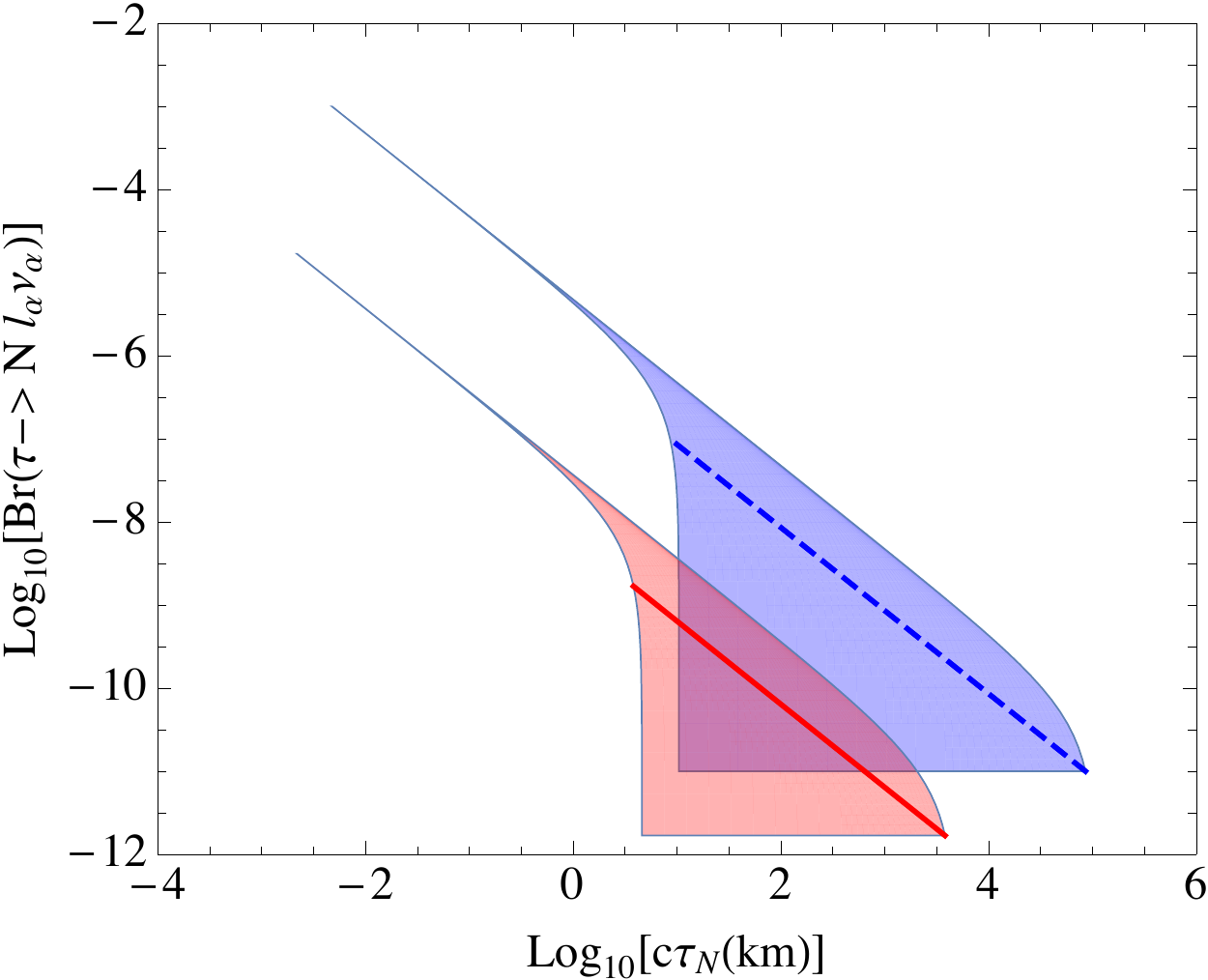}  
\end{center}
\caption{Allowed ranges for Br($D\rightarrow N$) (left panel) and  Br($D_{(s)}\rightarrow N$) (middle panel) and Br($\tau^\pm \rightarrow N l_\alpha^\pm \nu_\alpha$) (right panel) versus $c\tau_N$, for $m_N= 0.5$ GeV (dashed line, blue region) and 1 GeV (solid line, red region). The lines correspond to $10^{-10} \leq |U_e|^2=|U_\mu|^2=|U_\tau|^2 \leq 10^{-6}(10^{-7})$  for the lighter (heavier) mass. The ranges correspond to $10^{-10} \leq |U_e|^2=|U_\mu|^2 \leq 10^{-6}(10^{-7})$ and  $10^{-10} \leq |U_\tau|^2 \leq 10^{-2} (10^{-3})$. }
\label{fig:BrvsctauN}
\end{figure}

As already outlined, in this model both the production and decay rates (that is, the lifetime of the LLP) are controlled by the same set of parameters and therefore are highly correlated. Figures~\ref{fig:BrvsctauN}  show the regions allowed on the ${\rm Br}(P\rightarrow N)$ vs 
$c\tau_N$ plane (for $P = D, D_s, \tau$) allowing the mixing matrix elements to vary within the presently allowed regions \cite{Atre:2009rg}, for two values of the HNL mass. As can be seen from both figures, in this model it is possible to achieve values of the lifetime in the right ballpark needed to obtain a signal in neutrino detectors ($c\tau\sim\mathcal{O}(10)~$~km, after boosting to the lab frame), although at the price of very small production branching ratios. 
%%%%%%%%%%%%%%%%%%%
%%%%%%%%%%%%%%%%%%%

\begin{figure}
\begin{center}
\includegraphics[width=0.7\textwidth,keepaspectratio]{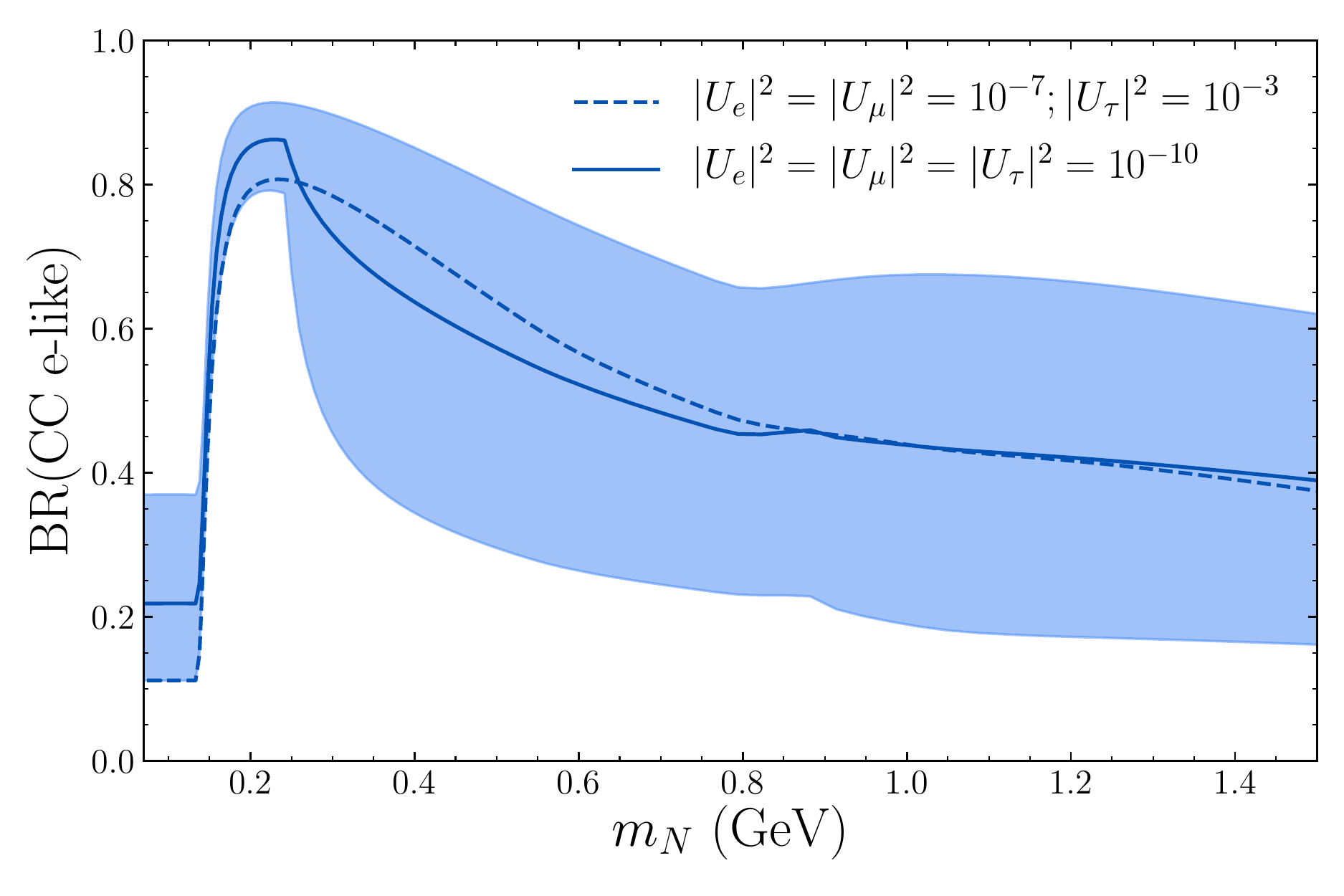} 
\end{center}
\caption{Value of the branching ratio of the HNL into e-like final states, defined in Eq.~\eqref{eq:CC-elike}, as a function of its mass. The lines correspond to fixed values of the mixing with the light neutrinos as indicated by the labels, while the shaded areas show the range of branching ratios accessible by varying independently the mixing matrix elements within the following limits: $10^{-10}< |U_{e}|^2 < 10^{-7}$, $10^{-10}< |U_{\mu}|^2 < 10^{-7} $ and $ 10^{-10}< |U_{\tau}|^2 < 10^{-4}$. For illustration purposes, we have chosen the same ranges for the mixing matrix elements as in Fig.~\ref{fig:BrvsctauN}. However, note that for specific mass ranges some of these values are excluded by laboratory constraints ~\cite{Artamonov:2014urb, CortinaGil:2017mqf, Abe:2019kgx, Mischke:2018qmv} (see also Refs.~\cite{Bryman:2019ssi, Atre:2009rg}).  }
\label{fig:BR-e-like}
\end{figure}

Finally, the number of events will also be proportional to the branching ratio of the decays leading to a cascade-like signature (at IC), or leading to e-like events (at SK). In practice, this amounts to adding the branching ratios into decay channels with either electrons, taus or hadronic resonances in the final state: 
\begin{eqnarray}
\label{eq:CC-elike}
{\rm Br (CC e-like)} & = & \sum_{M} \sum_{\alpha = e}{\rm Br}(N \to M^- l_\alpha^+) +
\sum_{M}  {\rm Br}(N \to M^0 \nu)  \nonumber \\
& + & \sum_{ \alpha=e,\tau} {\rm Br}(N \to \nu l_\alpha^- l_\alpha^+) + 
\sum_{\alpha, \beta=e,\tau} {\rm Br}(N \to \nu l_\alpha^- l_\beta^+) \, ,
\end{eqnarray}
where the processes involving light neutrinos include the diagrams for all active neutrino flavors, and the sum over $M$ includes all charged or neutral pseudoscalar and vector mesons below the mass of the $N$ ($\pi^\pm, K^\pm, \rho^\pm,\pi^0,\eta,\eta^\prime,\rho^0$, etc). The result is shown in Fig.~\ref{fig:BR-e-like} as a function of the mass of the HNL. 
%%%%%%%%%%%%%%%%%%%
%%%%%%%%%%%%%%%%%%%

\subsection{Dark Photons}
\label{sec:darkphoton}

Extensions of the Standard Model with an extra secluded $U(1)$ gauge boson are ubiquitous in the BSM landscape. We first consider a model where the dark photon $V$ is coupled to the visible sector via kinetic mixing~\cite{Holdom:1985ag}:
\begin{eqnarray}
 {\mathcal L}_V =  {\mathcal L}_{SM} -{1\over 4} V_{\mu\nu} V^{\mu\nu} + {\epsilon \over 2} B_{\mu\nu} V^{\mu\nu} - {1\over 2} m_V^2 V_\mu V^\mu \, ,
\label{eq:Vmodel}
\end{eqnarray}
where $B_{\mu\nu}$ is the hypercharge field strength tensor and $V_{\mu\nu}$ is analogously defined for the dark $U(1)$, which is assumed spontaneously broken. The Lagrangian in Eq.~\eqref{eq:Vmodel} implies that the dark photon couples universally to all charged particles, like the SM photon, although with a coupling that is reduced by the factor $\epsilon$. An enormous amount of work has been done in recent years to derive bounds on this scenario from beam dump experiments, $e^+ e^-$ and $pp$ colliders, neutrino scattering and other intensity frontier experiments. A recent summary of present bounds can be found in \cite{Bauer:2018onh}. 

\begin{figure}
\begin{center}
\includegraphics[width=0.8
\columnwidth]{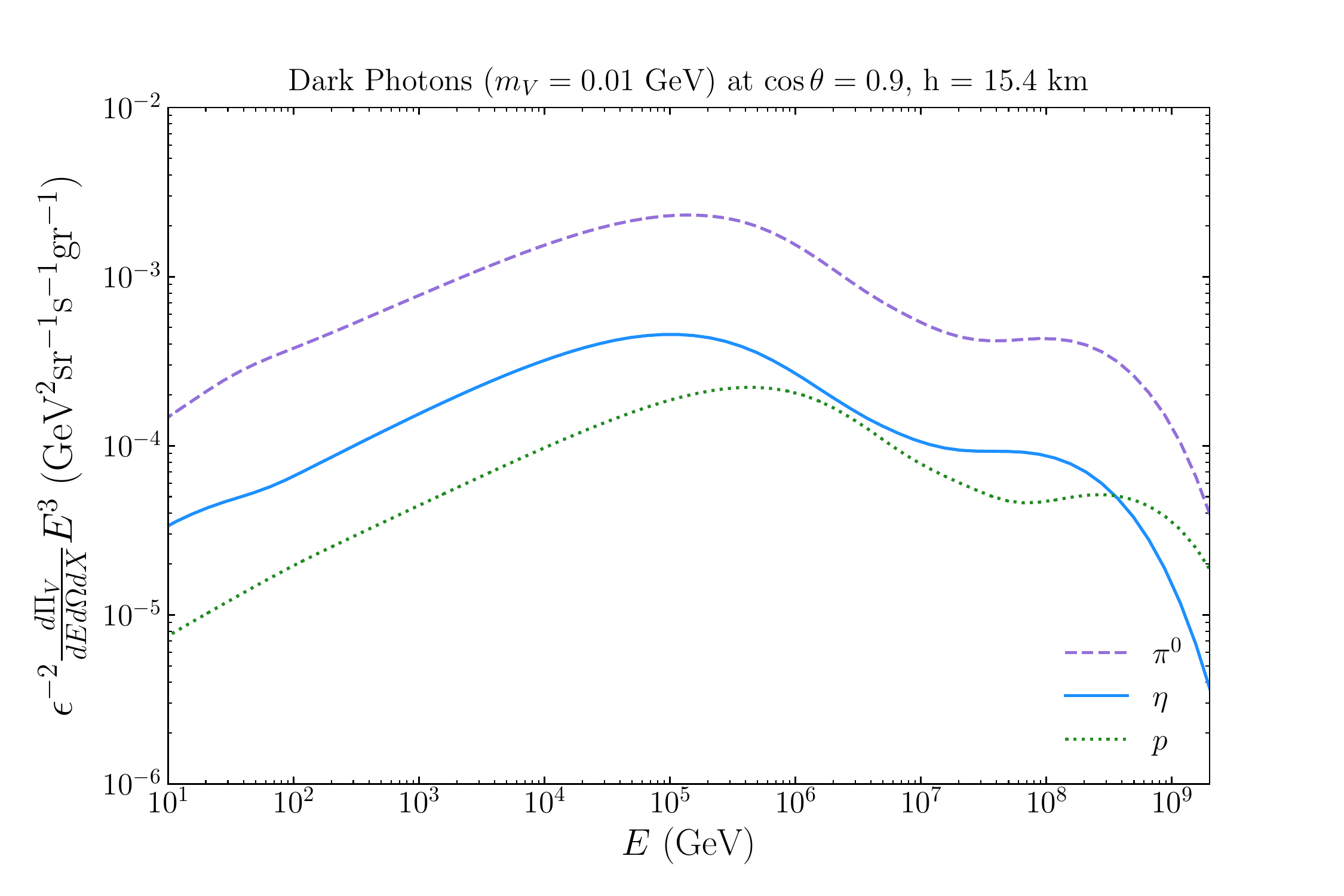} 
\end{center}
\caption{Production profile of dark photons produced from $\pi_0, \eta$ decay and proton \bremm, for $m_V=10^{-2}$~GeV. The profile is shown at approximately 15~km from the Earth's surface, since this is where the parent meson fluxes are expected to be maximal, see Fig.~\ref{fig:parentsh}, and for $\cos\theta = 0.9$ since we expect the signal to be largest as $\cos\theta \to 1$.
\label{fig:darkV}}
\end{figure}

The mechanisms leading to the production of dark photons in the atmosphere are the same as in a proton beam dump, which has been extensively studied in the literature, see \eg Refs.~\cite{Bjorken:2009mm,Essig:2010gu,Blumlein:2013cua}. 
For $m_V \leq m_{\pi_0}(m_{\eta'})$, the dominant production channel is the two-body decay $\pi_0(\eta') \rightarrow \gamma V$. For heavier masses and $m_V \leq m_p$, it can be produced via \bremm. The flux of $V$ from neutral meson decays can be obtained from Eqs.~(\ref{eq:master}), (\ref{eq:dn2b}) and (\ref{eq:2bd}), with the production branching ratios~\cite{Batell:2009di}:
\begin{eqnarray}
{\rm Br}(P \rightarrow V \gamma) \simeq \epsilon^2 \left( 1- {m_V^2\over  m_P^2}\right) {\rm Br}(P \rightarrow \gamma \gamma),
\end{eqnarray}
for $P= \pi_0, \eta$. 
The flux from \bremm is obtained from Eqs.~(\ref{eq:brem}) and (\ref{eq:bremdn}). The production profile of dark photons from these processes in units of $\epsilon^{-2}$ are shown in Fig.~\ref{fig:darkV}. 
%%%%%%%%%%%%%%%%
%%%%%%%%%%%%%%%%

\begin{figure}
\begin{center}
\includegraphics[width=0.7
\columnwidth]{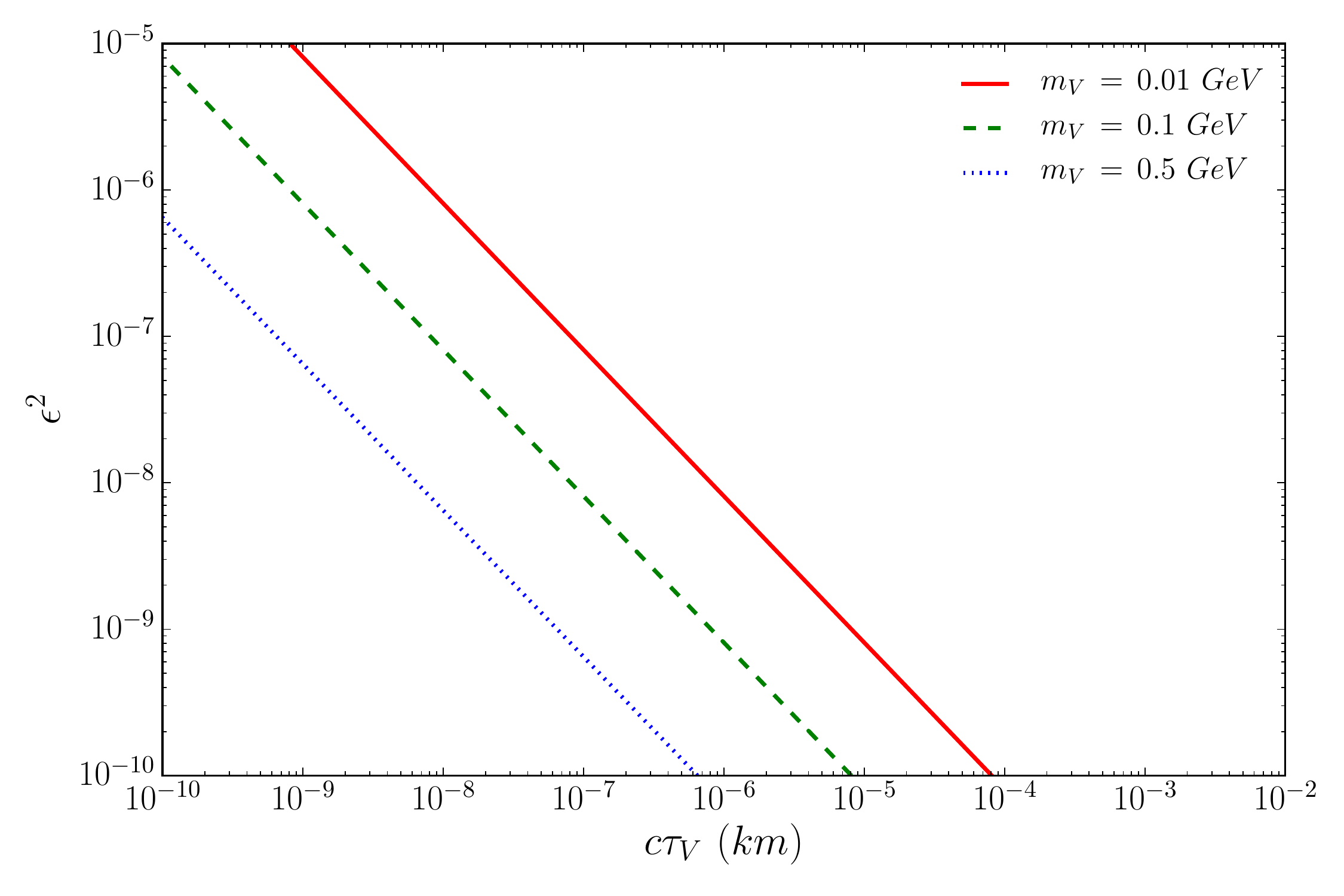} 
\end{center}
\caption{Correlation between the value of $\epsilon$ and the lifetime of the dark photon in its rest frame, $c\tau_V$, for three values of $m_V$: 0.01~GeV (solid), 0.1~GeV (dashed) and 0.5 GeV (dotted). }
\label{fig:evsctauV}
\end{figure}

The next ingredient we need in order to compute the expected number of decays in the detector is the lifetime of the dark photon. The partial decay widths into leptonic and hadronic channels can be found, for example, in Refs.~\cite{Bjorken:2009mm,Essig:2010gu,Blumlein:2013cua}:
\begin{eqnarray}
\Gamma( V \rightarrow l^+ l^-) = {1 \over 3} \epsilon^2 \alpha_{QED} m_{V}  \sqrt{1-{ {4 m_l^2\over m_{V}^2} }} \left(1+  {2 m_l^2\over m_{V}^2}\right), \\
\Gamma( V \rightarrow {\rm hadrons}) = {1 \over 3} \epsilon^2 \alpha_{QED} m_{V}  {\sigma(e^+ e^- \rightarrow {\rm hadrons})\over \sigma(e^+ e^- \rightarrow \mu^+ \mu^-)} \, ,
\end{eqnarray}
where $l$ stands for a SM charged lepton and $\alpha_{QED}\equiv e^2/(4\pi)$ stands for the SM fine structure constant. Adding up all possible decay channels below the hadronic threshold, and neglecting the electron mass, we get:
 \begin{equation}
 c\tau({\rm km})\simeq {8.1 \cdot 10^{-17}\over\epsilon^2} {{\rm GeV} \over m_V}  \left[1+ \sqrt{1 -4 {m_\mu^2\over m_V^2}} \left(1+ 2 {m_\mu^2\over m_V^2}\right)\right]^{-1}, \quad 2 m_\mu \leq m_V \leq 2 m_\pi \, .
 \end{equation}
Conversely, above the hadronic production threshold we can approximate the lifetime by the fitted formula:
 \begin{equation}
 c\tau({\rm km}) \approx {3 \cdot 10^{-17}  \over\epsilon^2} {{\rm GeV} \over m_V}  \left[ 1 + 2.54 {m_V\over {\rm GeV}} -4.76 \left({m_V\over {\rm GeV}}\right)^2\right], \quad 2 m_\pi \leq m_V \leq m_\rho.
  \end{equation} 
Again in this case, both the production rate and the decay length depend on just two  parameters: $\epsilon$ and $m_V$. Figure~\ref{fig:evsctauV} shows the correlation of $\epsilon$ and the lifetime of the dark photon, for three different values of $m_V$ in the range considered in this work.
%%%%%%%%%%%%%%%%%
%%%%%%%%%%%%%%%%%

\begin{figure}
\begin{center}
\includegraphics[width=0.7
\columnwidth]{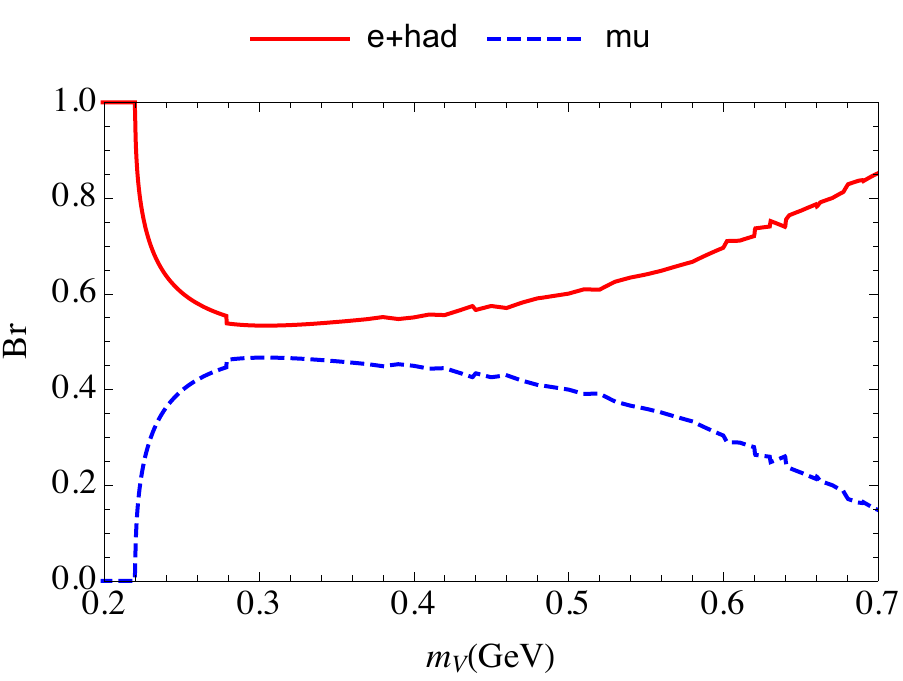} 
\end{center}
\caption{Branching ratios of the dark photon to electron/hadronic or muon decay channels as a function of the mass. }
\label{fig:brV}
\end{figure}

Finally, in our sensitivity calculations we focus on decays of the LLP leading to an electron-like signal in the detectors (\ie decays with electrons or hadrons in the final state). Therefore, in order to compute the number of events in this sample we need the corresponding branching ratio. The result is shown in Fig.~\ref{fig:brV} as a function of the mass of the dark photon.
%%%%%%%%%%%%%%%%
%%%%%%%%%%%%%%%%

\subsection{Heavy Neutral Leptons and $U(1)_{B-L}$ gauge symmetry}

We have finally considered an extension of the SM that includes both HNL and a new $U(1)$ gauge symmetry associated to the $B-L$ number. Models with HNL and extra $U(1)$ 
have been recently discussed in the context of the MiniBoone anomaly \cite{Bertuzzo:2018itn,Ballett:2018ynz,Arguelles:2018mtc,Coloma:2019qqj}, as well as in the context of explaining the matter-antimatter asymmetry or/and the observed dark matter abundance in the Universe~\cite{Caputo:2018zky}. The relevant terms in the Lagrangian are
\begin{eqnarray}
{\mathcal L} = {\mathcal L}_N -{1\over 4} V_{\mu\nu} V^{\mu\nu} -{1\over 2} m_V^2 V_\mu V^\mu + g_{\rm B-L} \left(\sum_f Q_{B-L}^f \overline{f} \gamma^\mu fV_\mu  - \overline{N} \gamma^\mu N V_\mu  \right), \label{eq:B-L}
\end{eqnarray}
where $g_{\rm B-L}$ is the gauge coupling of the dark photon, and the sum runs over all standard model fermions, which are charged under this new symmetry ($Q_{B-L}^f = 1/3$ for quarks, and $Q_{B-L}^f = -1$ for leptons). In this model, the origin of the B-L gauge boson mass typically requires a more complex hidden sector. However, our approach is phenomenological and therefore we will not address the possible origin of the the mass term which this is beyond the scope of this work.
Finally, note that in principle a kinetic mixing term could also be added to Eq.~\eqref{eq:B-L}. If the kinetic mixing is large, the phenomenological consequences of the model reduce to the dark photon case, which was already discussed in Sec.~\ref{sec:darkphoton}. Here we focus on the opposite limit instead, when the kinetic mixing term is negligible and the coupling dominates, since it leads to a different phenomenology. While a kinetic mixing term is always generated at loop level, this is expected to be subleading. 

An interesting feature of this model is that the production of the $N$ is dominated by the decays of the dark photon $V\rightarrow NN$, provided $m_N \leq m_V/2$, while its decay is controlled by the mixing with the light neutrinos. We might expect in this case to have an enhanced production if $g_{B-L}$ is not too small, while the $N$ might be very long-lived as shown in Fig.~\ref{fig:BrvsctauN}. 
Present upper limits \cite{Ilten:2018crw,Bauer:2018onh,Heeck:2018nzc} on $g_{B-L}^2$ are around $\sim 10^{-7}$ in the mass range of interest, around $m_V \sim {\mathcal O}$(GeV). From Fig.~\ref{fig:BrvsctauN} we see that  $c\tau_N$ in the range $[10^{-2},10^5]$~km for $m_V \sim 0.5$~GeV are allowed.

\begin{figure}
\begin{center}
\includegraphics[width=0.7
\columnwidth]{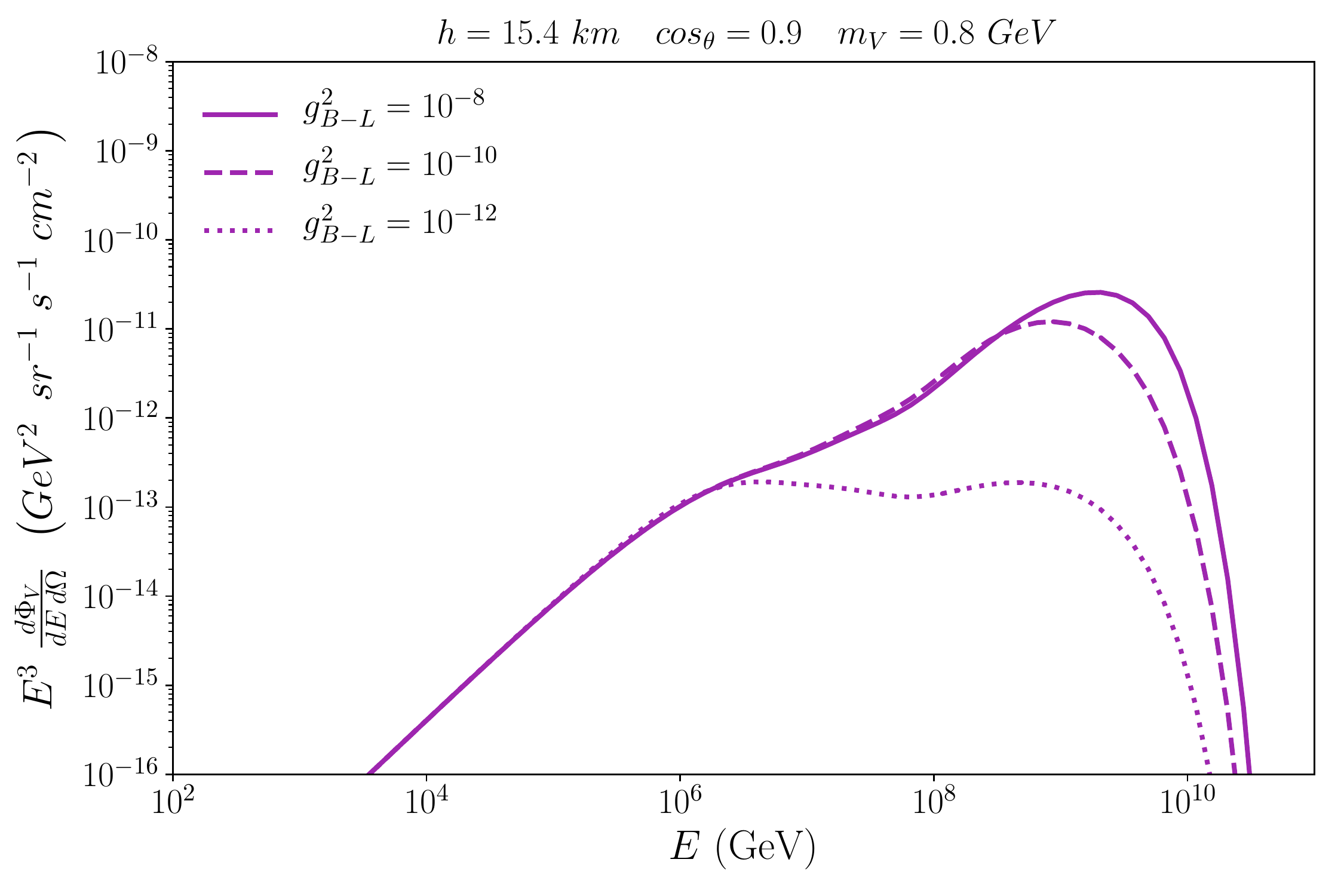} 
\end{center}
\caption{$V$ fluxes in the $B-L$ model with $m_V=0.8$GeV and for three values of $g_{B-L}^2$. }
\label{fig:fluxesVBmL}
\end{figure}

Neglecting the decay of the dark photon, the flux of $V$ can be obtained as in Sec.~\ref{sec:darkphoton} simply replacing~\cite{Bauer:2018onh}:
\begin{eqnarray}
\alpha_{QED} \epsilon^2 \rightarrow \alpha_{B-L} \equiv \frac{g_{\rm B-L}^2}{4\pi} \, .
\end{eqnarray}
However, in the $B-L$ model the dark photon decays very promptly. Its lifetime (for $2 m_N \leq m_V $, and $2 m_\mu \leq m_V \leq m_\omega$) is approximately given by
 \begin{equation}
 c\tau_V({\rm km})\simeq {0.74 \cdot 10^{-17}\over g_{B-L}^2} {{\rm GeV} \over m_V}  \left[{5\over 2}+ \sqrt{1 -4 {m_\mu^2\over m_V^2}} \left(1+ 2 {m_\mu^2\over m_V^2}\right)+{1 \over 2} \left(1 - {4 m_N^2\over  m_V^2}\right)^{3/2} \right]^{-1} \, .
 \end{equation}
Therefore, decay effects should be accounted for in the calculation of the flux, following the same arguments as in App.~\ref{app:fluxes}. It is easy to show that, if we denote by $\tilde\phi_V$ the dark photon flux obtained without accounting for the decay, then the solution to the cascade equation including decay effects is given by:
\begin{equation}
\label{eq:Vflux-decay}
\phi_V(\ell)= \int_\ell^{\infty } d\ell^\prime e^{- {\ell^\prime - \ell\over \ell_{\rm decay}} }~{ d\tilde\phi_V\over d \ell^\prime }, 
\end{equation}
where $\ell_{\rm decay} = \gamma_V \beta_V c \tau_V$. 
The flux of dark photons taking into account the decay is shown in Fig.~\ref{fig:fluxesVBmL} for a three different values of $g_{B-L}$ in the experimentally allowed range. 
%%%%%%%%%%%%%%%%%%
%%%%%%%%%%%%%%%%%%
Interestingly, while at high energies there is clearly a suppression of the dark photon flux, which depends on the value of $g_{\rm B-L}$, at low energies the production remains practically independent from the value of $g_{B-L}$. This can be understood as follows. In this part of the spectrum the lifetime of the dark photon is very short and the exponential in Eq.~\eqref{eq:Vflux-decay} goes to zero very rapidly, except for values of $\ell^\prime$ such that $\Delta \ell = \ell^\prime - \ell \propto c\tau_V$. Thus, at low energies the resulting flux will take the form
\begin{equation}
\label{eq:Vflux-approx}
\phi_V \propto c\tau_V \frac{d\tilde\phi_V}{d \ell } \, ,
\end{equation}
and since $c\tau_V \propto g_{\rm B-L}^{-2}$ this effectively renders the result independent of $g_{\rm B-L}$ at low energies.

From these fluxes, we can get the production profile of $N$ as in Eq.~(\ref{eq:master}), where now the parent particle $P$ is the dark photon instead of SM mesons. The decay $V\rightarrow N N$ is two-body and therefore the decay distribution is the same as in Eqs.~(\ref{eq:dn2b}) and (\ref{eq:2bd}). Neglecting the electron mass, its partial decay width reads:
  \begin{eqnarray}
  \Gamma(V\rightarrow NN) \simeq {1 \over 2} \left(1 -{4 m_N^2\over m_V^2}\right)^{3/2}  \Gamma(V\rightarrow e^+e^-) \, .
  \end{eqnarray} 
Finally, the branching ratio in this channel (in the mass region of interest) can be well approximated by:
 \begin{eqnarray}
{ \rm Br}(V\rightarrow N N) \simeq {1 \over 2} \left(1 -{4 m_N^2\over m_V^2}\right)^{3/2}   \left[{5\over 2}+ \sqrt{1 -4 {m_\mu^2\over m_V^2}} \left(1+ 2 {m_\mu^2\over m_V^2}\right) +{1 \over 2} \left(1 -{4 m_N^2\over m_V^2}\right)^{3/2} \right]^{-1}, \nonumber\\
\end{eqnarray}
since the decays into mesons are suppressed. 

\begin{figure}
\begin{center}
\includegraphics[width=0.7
\columnwidth]{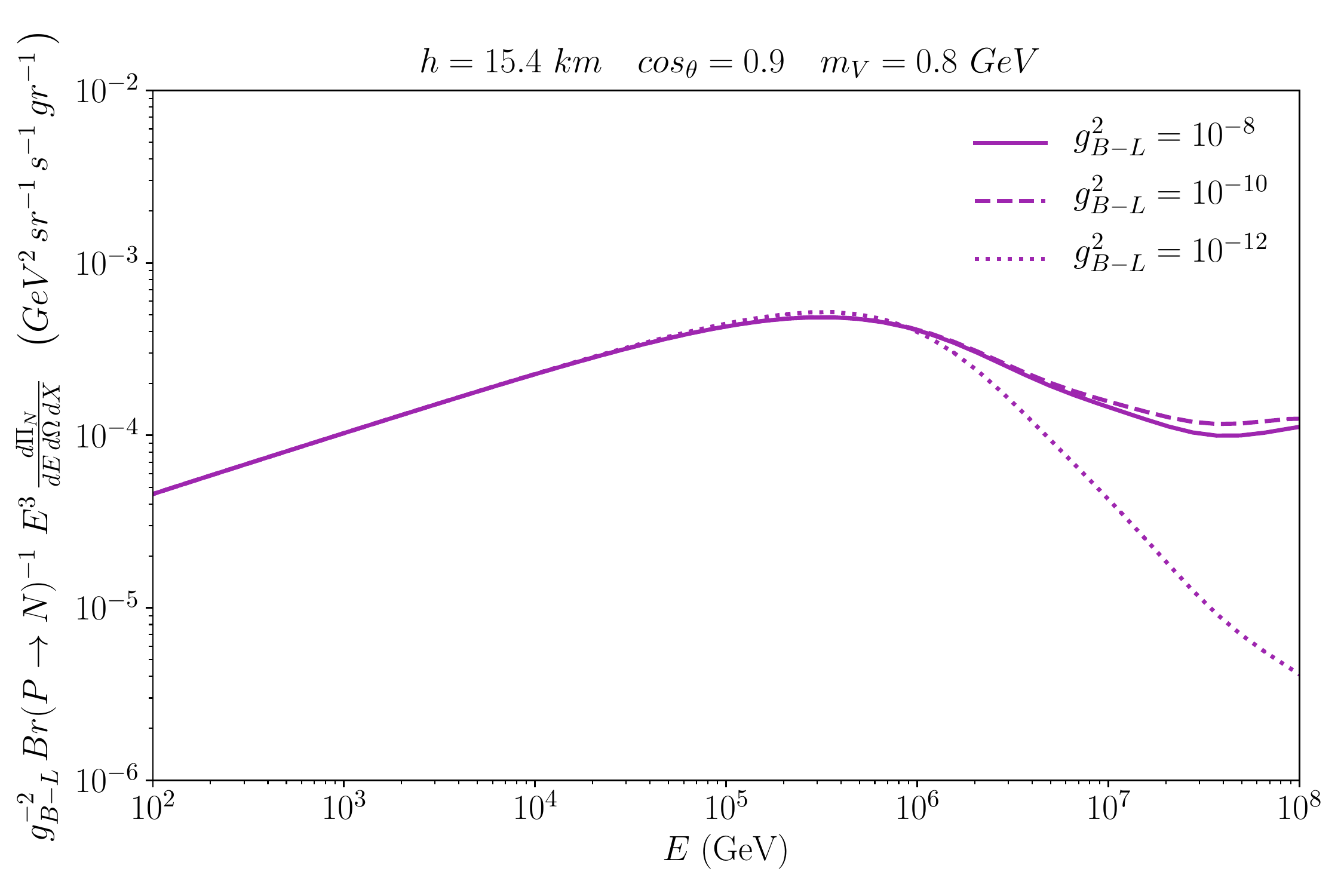} 
\end{center}
\caption{$N$ production from $V$ decays  in the $B-L$ model with $m_V=0.8$GeV, $m_N=0.35$ GeV and for three values of $g_{B-L}^2$. Note that because two $N$ are
produced in each decay we define ${\rm Br}(V\rightarrow N) = 2 {\rm Br}(V\rightarrow NN).$ }
\label{fig:fluxesNBmL}
\end{figure}

The resulting production profile of $N$ from dark photon decays is shown in Fig.~\ref{fig:fluxesNBmL}. Since the production and decay processes are effectively decoupled, the decay length of the $N$ (which is controlled by the mixing) can now take a wide range of values (as shown in Fig.~\ref{fig:BrvsctauN}) without affecting the production rates.
%%%%%%%%%%%%%%%%%%%%%%%
%%%%%%%%%%%%%%%%%%%%%%%

\section{Results}
\label{sec:results}

In this section we present our numerical results for the sensitivity to the scenarios described in Sec.~\ref{sec:LLP}. Using the data sets described in Sec.~\ref{sec:detect}, a $\chi^2$ fit to the data is performed. A Poissonian $\chi^2$ function has been used, defined as:
\begin{eqnarray}
\chi^2({\rm LLP}) =2  \sum_{i, \alpha}  \left(N_i^{\alpha}+ B_i^{\alpha} - n_i^{\alpha}+n_i^{\alpha} \log\left({n_i^{\alpha}\over N_i^{\alpha}+ B_i^{\alpha}}\right)\right),
\end{eqnarray}
where the sum runs over the energy (angular) bins for the IC (SK) detector analyses, and we use $\alpha=\rm{cascade}$(e-like) for IC (SK). Here, $n_i^\alpha$ stands for the data observed in each bin while $N_i^\alpha$ is the predicted number of signal events and $B_i^\alpha$ is the background prediction, which includes the predicted number of atmospheric neutrino events in the SM (plus the contribution from the fitted astrophysical neutrino flux, in the case of IC). 

A comment regarding the treatment of the astrophysical neutrino background for IC is in order, since the distribution of astrophysical neutrinos is obtained from a fit to the same data that we use in order to derive a limit to LLP models. Although this may seem inconsistent, it should be noted that due to the nature of the decay and the spectra of the parent mesons in the atmosphere (which obey a power law), at IC we expect most of our sensitivity to come from events observed at energies below 10~TeV, where the contribution from the astrophysical neutrinos is subdominant. Moreover, from the comparison between the left and right panels in Fig.~8 in Ref.~\cite{Aartsen:2014muf}, it seems that the fit to the astrophysical neutrino background is mostly driven by the data from the northern sky, whereas we expect LLP decays to contribute mostly to the southern sky data set, which is the only data we use. Therefore, although a more detailed analysis by the experimental collaboration would be necessary to do this analysis properly, we believe the results would not be very different. Of course, given the very characteristic zenith angle dependence of the LLP signal, a greater sensitivity is expected  if 2D binned data (using both energy and angular information) were to be used instead. 

\begin{figure}
\begin{center}
\includegraphics[width=0.48
\columnwidth]{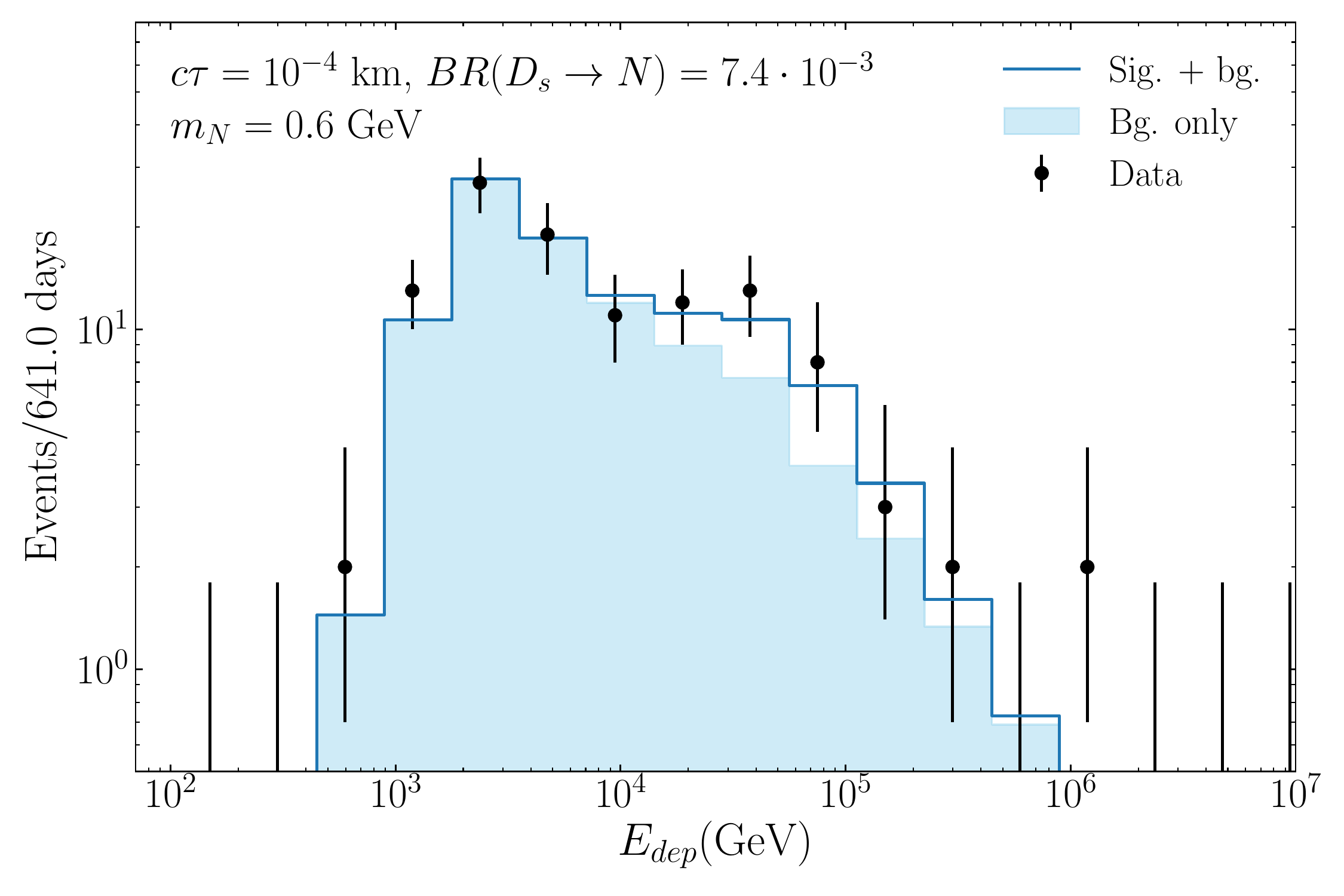} 
\includegraphics[width=0.48
\columnwidth]{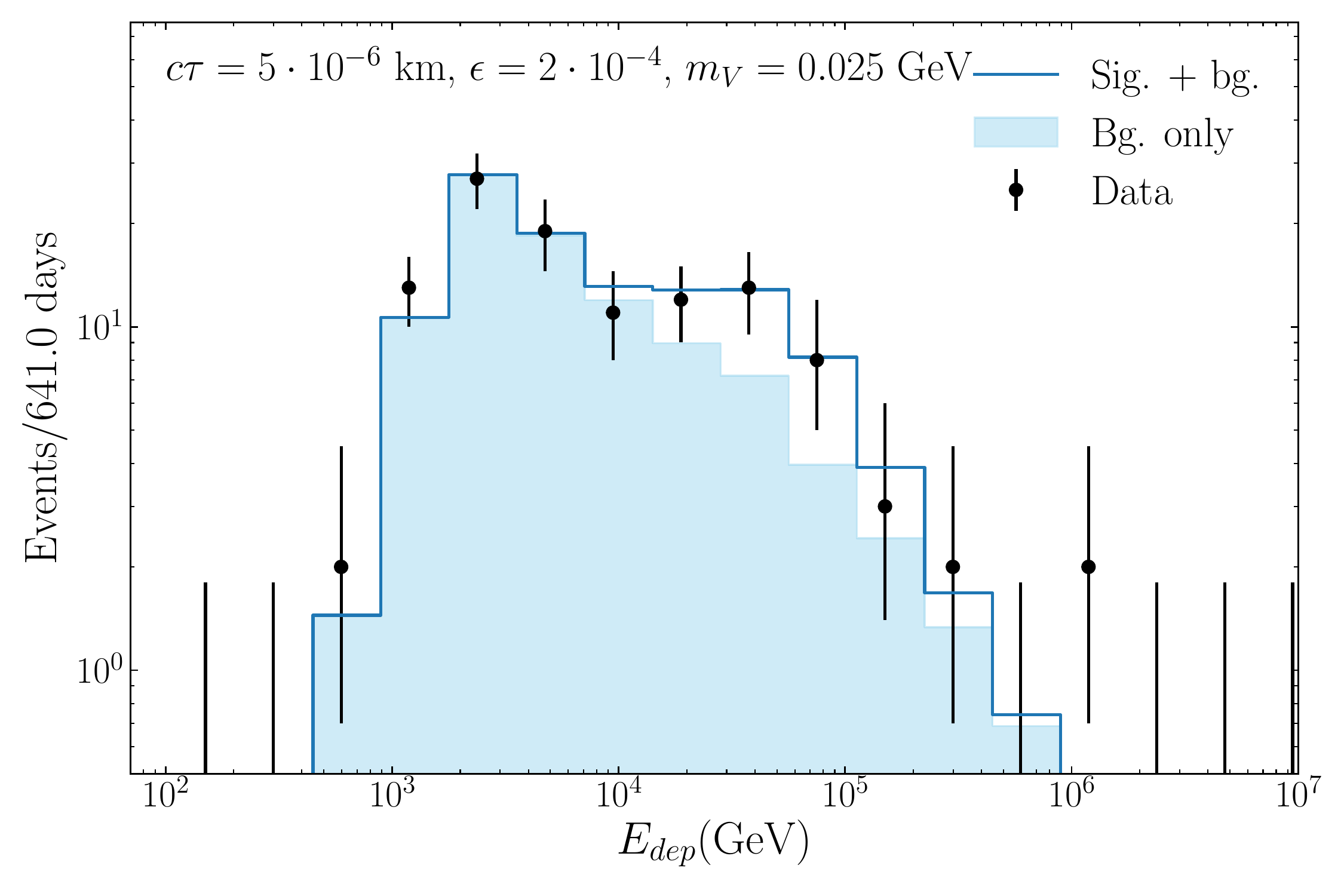}  
\end{center}
\caption{ Expected number of decays for the IC Medium Energy Starting Events (MESE) sample, as a function of the deposited energy in the detector and for the down-going sample only ($\cos\theta > 0$), for two of the scenarios considered in this work. The shaded histograms show the background prediction, including a fitted distribution to the astrophysical neutrino signal, the cosmic muon background, and atmospheric neutrino events. The black dots correspond to the observed data with error bars, as in Ref.~\cite{Aartsen:2014muf}. In the left panel, the signal has been computed for a HNL with a mass $m_N=0.6$~GeV, a $c\tau_N=  10^{-4}$~km and ${\rm Br}(D_s \rightarrow N)$ Br($N\rightarrow$ CC e-like) $=7.4\cdot 10^{-3}$. In the right panel, the signal has been computed for a dark photon with a mass $m_V=25$~MeV for $\epsilon =2\cdot 10^{-4}$ and  an (uncorrelated) lifetime $c\tau_V=  5 \cdot10^{-6}$~km.
}
\label{fig:data-ic}
\end{figure}

For illustration, Fig.~\ref{fig:data-ic} shows the expected number of signal events for HNL (left panel) and dark photons (right panel), for an assumed value of the production branching ratio of the HNL and its lifetime. The shaded histograms show the background prediction, while the solid lines show the expected signal plus background event rates per bin. For reference, the black points show the observed data, as given in Ref.~\cite{Aartsen:2014muf}.
%%%%%%%%%%%%%%%%%%%%%%%%%%
%%%%%%%%%%%%%%%%%%%%%%%%%%

\begin{figure}
\begin{center}
\includegraphics[width=0.7
\columnwidth]{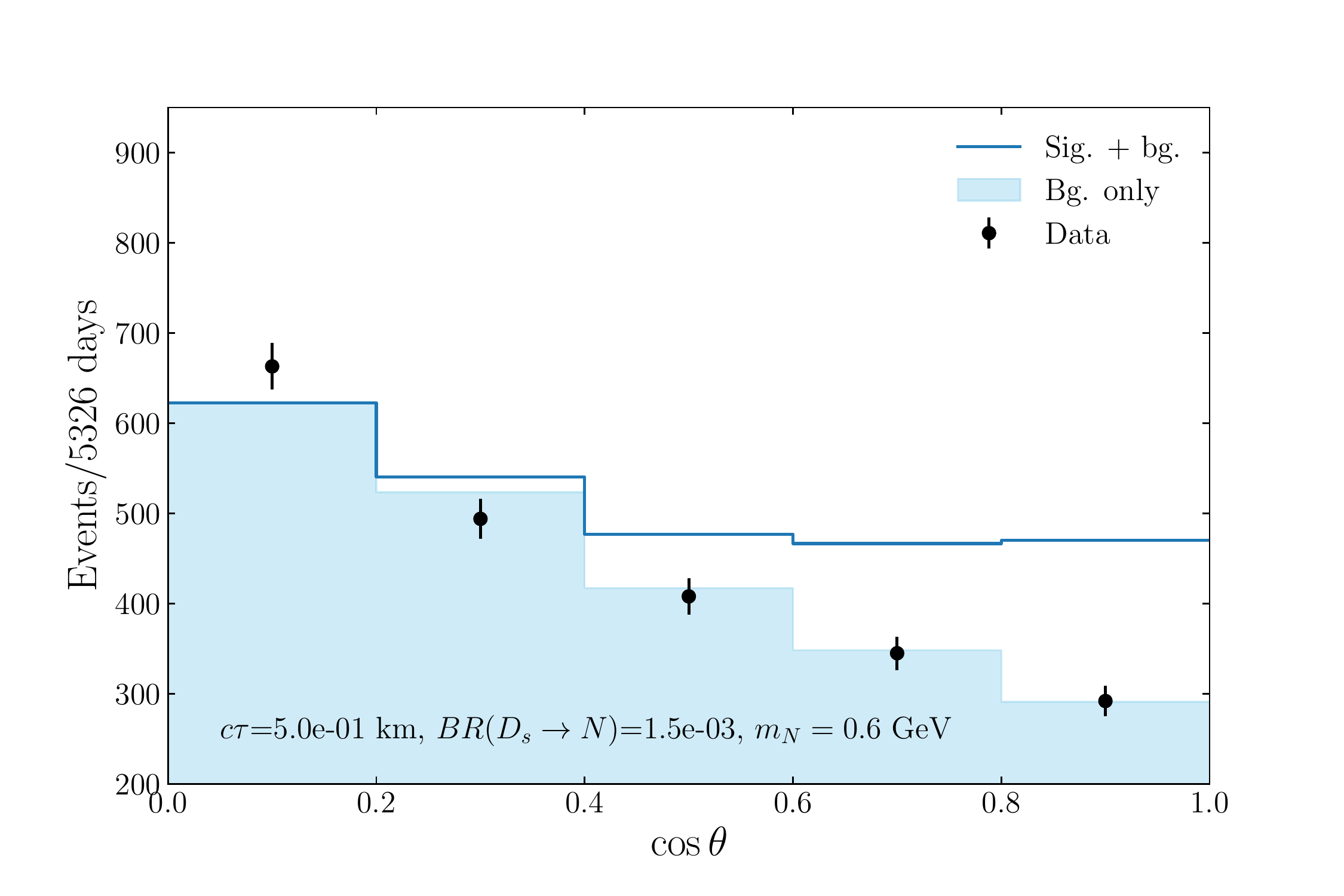} 
\end{center}
\caption{ Expected number of decays in SK, as a function of the zenith angle. The shaded histograms show the atmospheric background prediction, while the solid histograms show the signal plus neutrino background. The signal in this case has been computed for a HNL with a mass $m_N=0.6$~GeV, a $c\tau_N=  0.5$~km and $\rm{Br}(D_s \rightarrow N) {\rm Br}(N\rightarrow {\rm CC e-like }) =1.5\cdot 10^{-3}$. The data points, extracted from Ref.~\cite{Abe:2017aap}, are shown with statistical error bars for reference.  }
\label{fig:eventsSK}
\end{figure}

The zenith angle distribution for the neutrino background and LLP is signal events in Super-Kamiokande is shown in Fig.~\ref{fig:eventsSK}. As expected the signal is peaked at small zenith angles, that is, most of the decaying LLP are expected to come from trajectories entering the detector from above.

Before moving on to the discussion of our results, we should note that no systematic uncertainties have been included in our calculations, since this is an extremely challenging task using only publicly available information. Eventually, a detailed study including a proper implementation of systematic uncertainties and detector reconstruction effects should be performed by the experimental collaborations themselves.  

\subsection{Exclusion limits for IceCube and Super-Kamiokande}

\begin{figure}
\begin{center}
\includegraphics[width=1.0
\columnwidth]{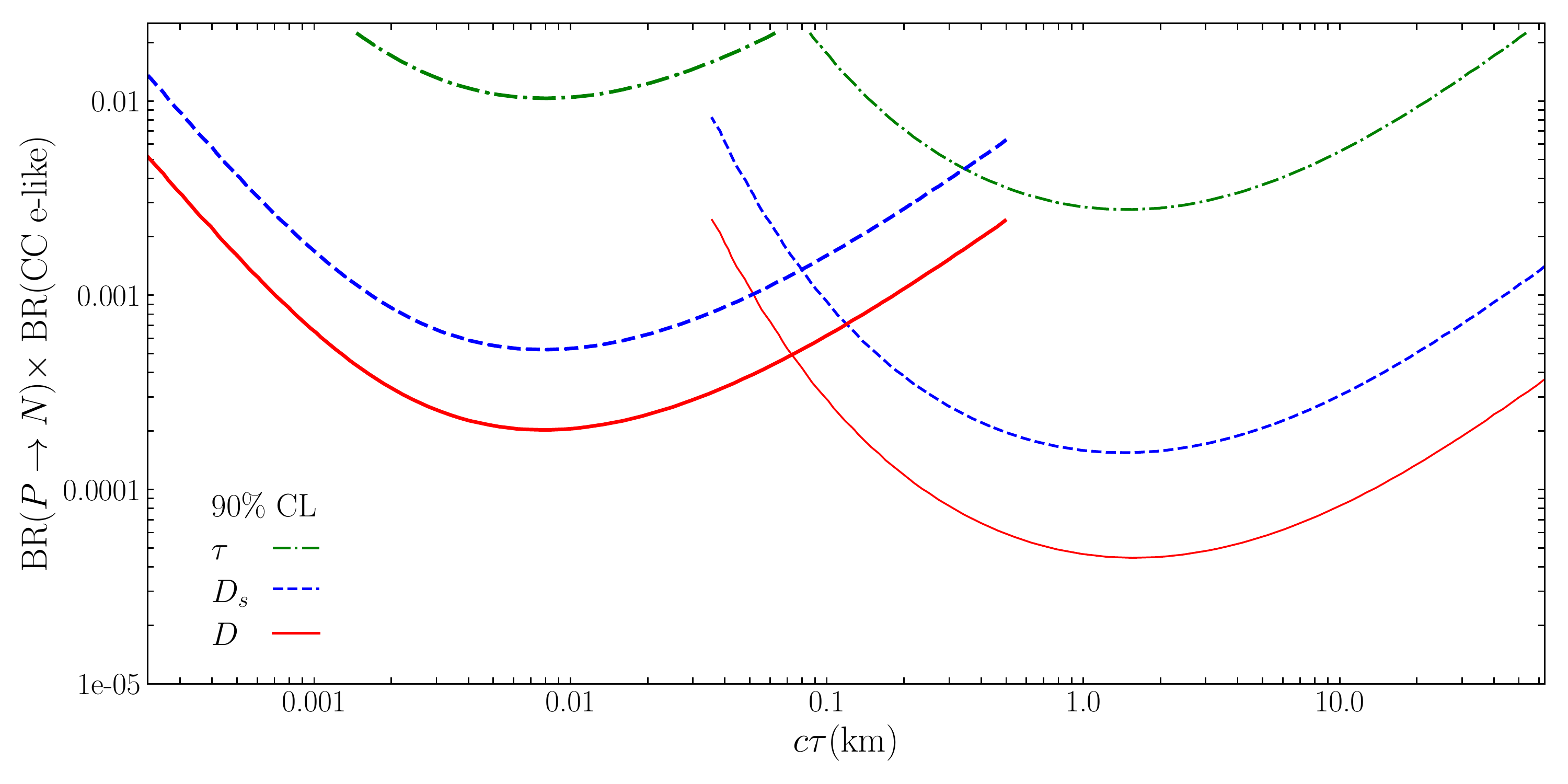} 
\end{center}
\caption{Limits  on HNL of $m_V=1$GeV  from IceCube (thick lines) and SuperKamiokande (thin lines) on the $BR(P\rightarrow N) \times BR(N\rightarrow$ CC-e~like) vs $c\tau$ plane including production from the parent particles  $P=D$ (solid) , $D_s$ (dashed) and $\tau$ (dash-dotted).  }
\label{fig:limitHNL}
\end{figure}

\begin{figure}
\begin{center}
\includegraphics[width=1.0
\columnwidth]{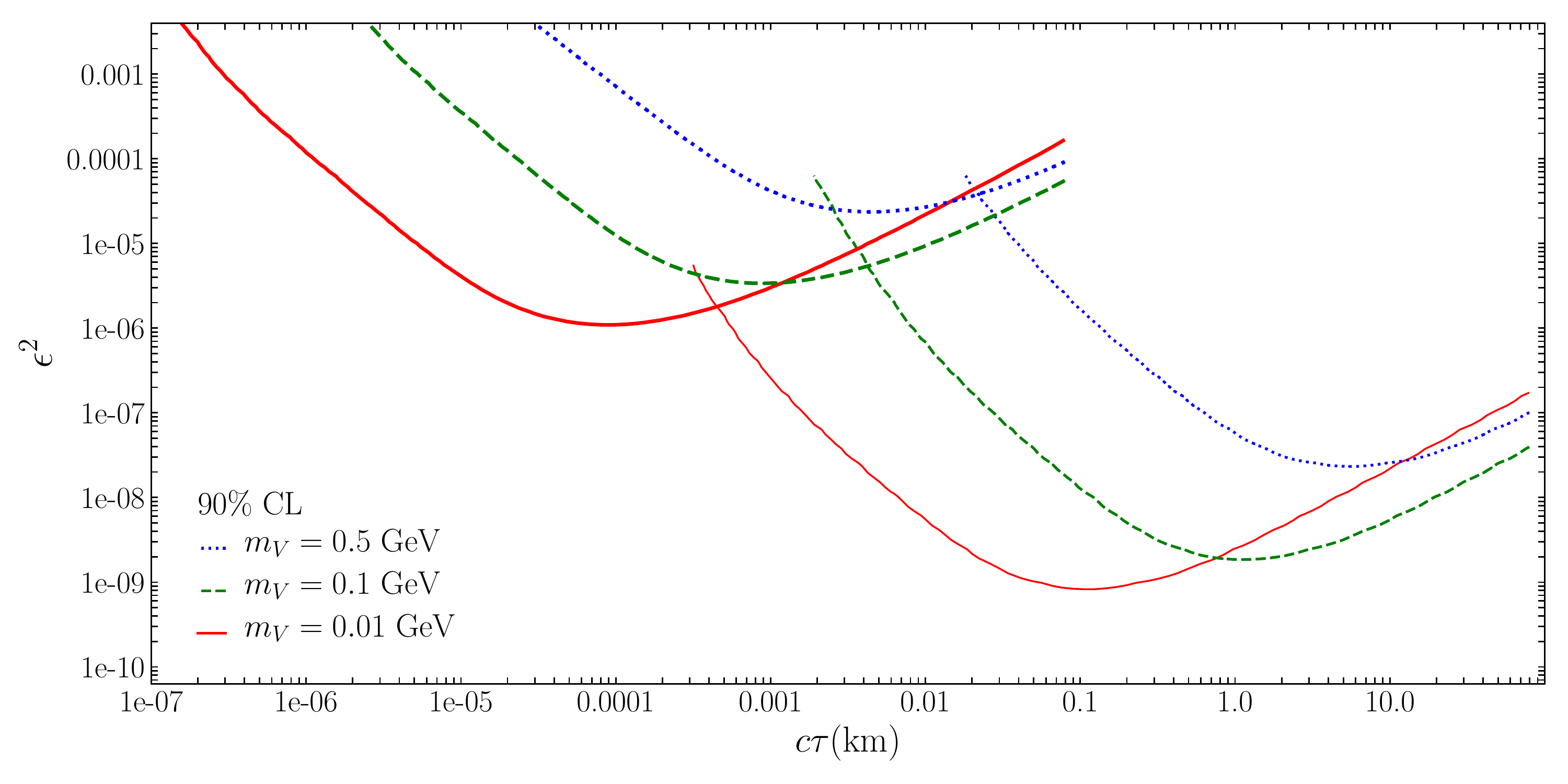} 
\end{center}
\caption{Limits  on dark photons decays from IceCube (thick lines) and Super-Kamiokande (thin lines) on the $\epsilon^2$ vs $c\tau_V$ plane including production from $\pi_0, \eta$ decay and \bremm for $m_V=0.01$GeV (solid), $0.1$GeV (dashed) and $0.5$ GeV (dotted). }
\label{fig:limitV}
\end{figure}

Following the procedure described above, we proceed to derive exclusion limits for the HNL and dark photon LLPs, for both IC and SK. 
In doing so, we consider that the data is chi-squared distributed with $n$ degrees of freedom (d.o.f.), $n$ being the number of bins in the data (that is, 5 bins in $\cos\theta$ in the case of SK, as opposed to 15 energy bins in the case of IC). Our exclusion limits thus indicate the region in parameter space where the $\chi^2$ value exceeds the corresponding one at 90\%~CL, regardless of where the best-fit point lies in the parameter space.  Our results are shown in Fig.~\ref{fig:limitHNL} for the HNL, assuming $m_N=1$~GeV, and in Fig.~\ref{fig:limitV} for the dark photons, for three different values of $m_V$ as indicated in the legend. In both figures, thick lines correspond to the IC limits, while thin lines indicate the SK result. 

For both IC and SK, the best limits are reached when the decay length in the lab frame is $c\tau_{\rm lab}\sim \mathcal{O}(10)$~km, as expected from naive arguments. However, the two experiments observe events in very different energy regimes: $\mathcal{O}({\rm TeV})$ in the case of IC, while $\mathcal{O}({\rm GeV})$ in SK, and therefore the boost factor (which also depends on the mass of the LLP) will be very different. As a result, their sensitivities are highly complementary and explore different regions of $c\tau$.
For IC, the best limit is reached for $c\tau \sim {\mathcal O}(10^{-3}-10^{-2})$~km in the case of LLP with masses $m \sim 0.5-1$~GeV, while for SK the best limits are obtained for $c\tau \sim {\mathcal O}(1-10)$~km. 

%%%%%%%%%%%%%%%%%%%%%%%
%%%%%%%%%%%%%%%%%%%%%%%
In the case of the HNL (Fig~\ref{fig:limitHNL}), the solid red, dashed blue, and dot-dashed green lines show our results assuming that the LLP is mainly  produced from $D$, $D_s$ or $\tau$ decays. Although our results are shown for $m_N=1$GeV, the regions do not change significantly for other values of $m_N$ between 0.5 and 1.5~GeV. As can be seen from this figure, the limits obtained for the two experiments are very similar, although slightly better for SK. However, in both cases the limits cannot probe the allowed regions of parameter space once we take into account the correlation between production and decay, as shown in Figs.~\ref{fig:BrvsctauN}. Therefore we conclude that, in the minimal scenario described in Sec.~\ref{sec:HNL}, these limits are not competitive (although this may not be the case in non-minimal BSM scenarios where the production and decay may be uncorrelated).

%%%%%%%%%%%%%%%%%%%%%%%
%%%%%%%%%%%%%%%%%%%%%%%
%
The exclusion limits for the dark photon scenario (Sec.~\ref{sec:darkphoton}) are shown in Fig.~\ref{fig:limitV}. In this case our results include all production mechanisms kinematically available for the production of the dark photon: $\pi^0$ decays, $\eta$ decays, and $p$ \bremm. The solid red, dashed green and dotted blue lines indicate the results obtained for three different values of the dark photon mass, as indicated in the legend. Again in this case, as for the HNL scenario, we find that this search cannot probe the region of parameter space shown in Fig.~\ref{fig:evsctauV} for correlated production and decay rates. However, the limits might be useful in the context of more complex models where production and decay are uncorrelated. In particular, it is noticeable that the limits for SK are three orders of magnitude better than for IC. This is because, in the case of light mesons and protons, the flux follows a harder power law than for $D_{(s)}$ mesons and taus, and therefore reducing the energy of the events detected leads to huge enhancement in the expected number of events.

%%%%%%%%%%%%%%%%%%%%%%%
\begin{figure}
\begin{center}
\includegraphics[width=0.7
\columnwidth]{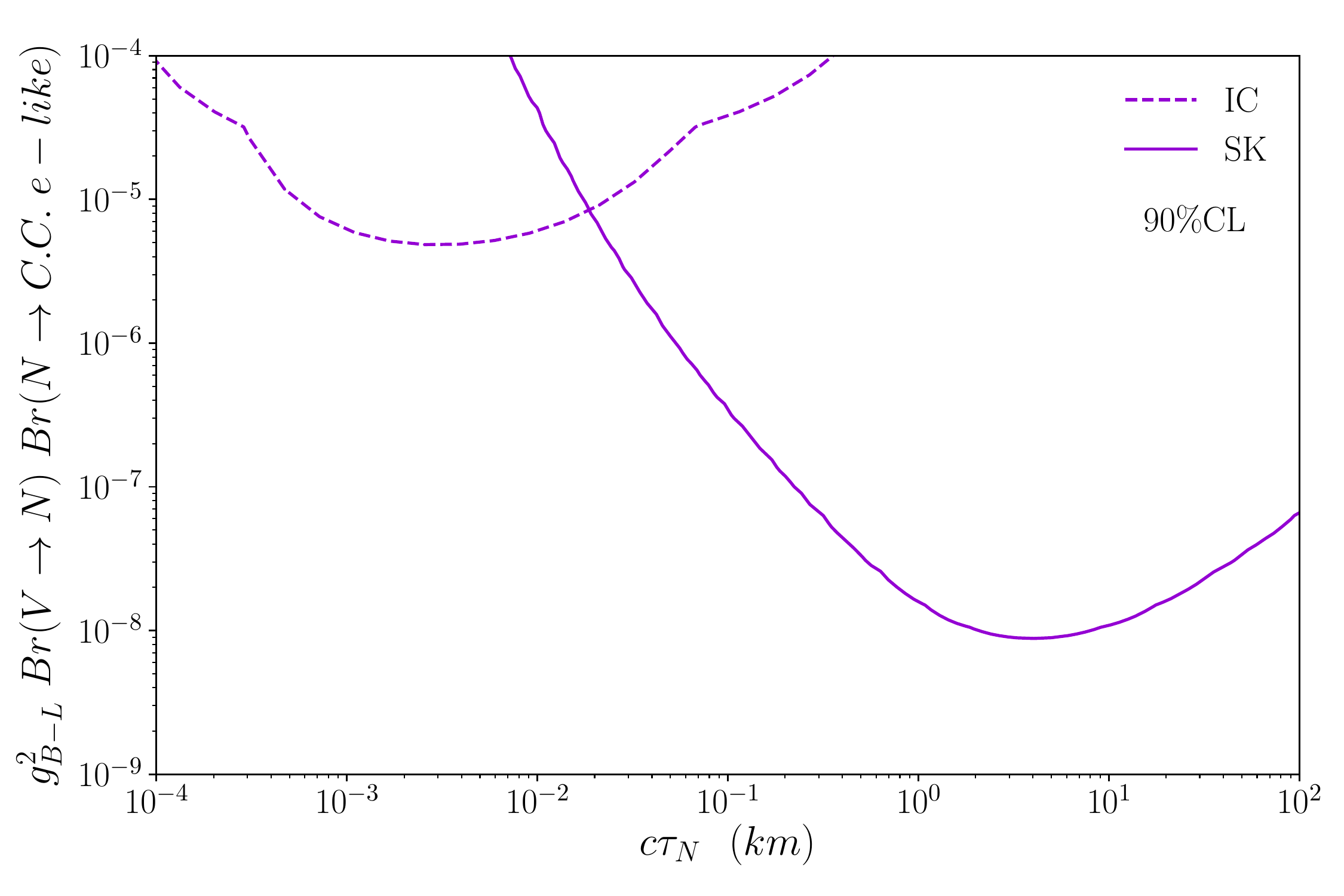} 
\end{center}
\caption{Limits  on dark photons decays from IceCube (thick lines) and Super-Kamiokande (thin lines) on the $g_{B-L}^2 {\rm Br}(V\rightarrow N)$ vs $c\tau_N$ for $m_V=0.8$GeV and $m_N=0.35$GeV, where ${\rm Br}(V \rightarrow N) = 2 {\rm Br}(V\rightarrow NN)$.}
\label{fig:limitHNLBL}
\end{figure}
%%%%%%%%%%%%%%%%%%%%%%%
%
On the other hand, in the $B-L$ model it is possible to uncorrelate production and decay since the LLP is not the 
dark photon, but the HNL to which it decays. Therefore, production rates in this model are controlled by the $B-L$ gauge coupling, while the decay depends on the mixing with the light neutrinos. The limits for the $B-L$ model on the plane $g_{B-L}^2 {\rm Br}(V\rightarrow N) {\rm Br}(N\rightarrow CC e-like)$ vs. $c\tau_N$  are shown in Fig.~\ref{fig:limitHNLBL}, for $m_V = 0.8$~GeV and $m_N=0.35$~GeV. As in the case of the dark photon, the limits for SK are three orders of magnitude better since the production in this case is controlled by $p$ \bremm . While IC limits fall short of competitiveness, assuming large enough branching ratios ${\rm Br}(V\to N) {\rm Br (CC e-like)}$ our results show that SK would be able to reach values of $g_{B-L}^2 \leq 10^{-7}$, which are comparable to BaBar limits in this region~\cite{Ilten:2018crw,Bauer:2018onh,Heeck:2018nzc}. At the same time, the values of $c\tau_N \sim {\mathcal O}(1)$~km needed for the HNL are allowed by present constraints, see Fig.~\ref{fig:BrvsctauN}. 

\subsection{IceCube preferred regions}

An interesting observation is that, in the case of IC,  the minimization of the $\chi^2$ on these two-dimensional planes gives a minimum of the $\chi^2$ that lies at a non-zero value of the LLP production rate. This is the result from a small excess in the data at around 30~TeV, as can be seen by naked eye in Fig.~\ref{fig:data-ic}. Although the source of the excess is unclear, and it might be related to an underestimation of the $\mu$ background or an unidentified source of systematic uncertainties, in the remainder of this section we explore possible explanations in the LLP models discussed in Sec.~\ref{sec:LLP}.

\begin{figure}
\begin{center}
\includegraphics[width=0.7
\columnwidth]{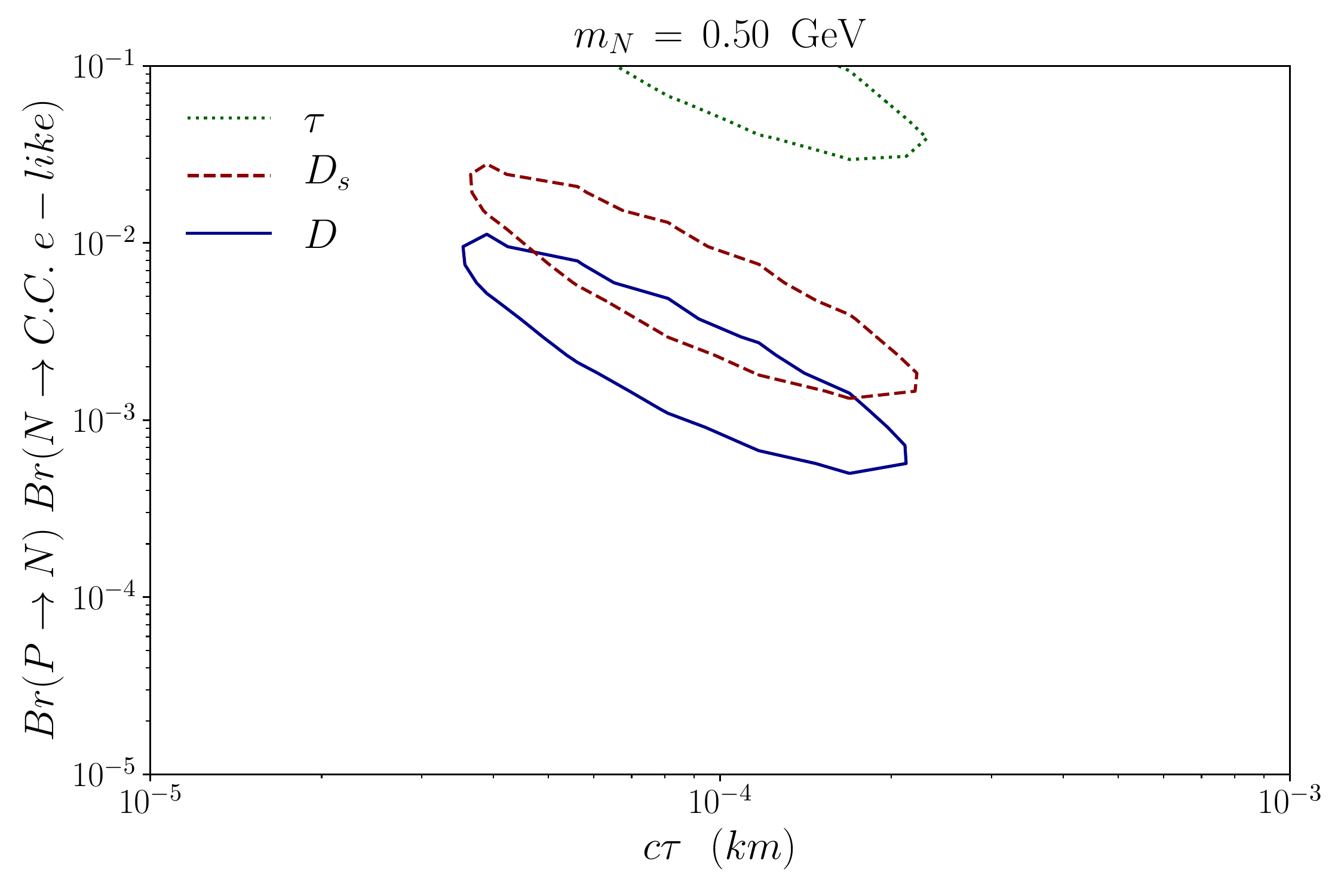} 
\end{center}
\caption{1$\sigma$ best-fit regions for HNL decays in IceCube assuming uncorrelated  Br($P \rightarrow N$) vs $c\tau$ for $m_N=0.5$~GeV and considering separately the production from three parent particles $P=D,D_s, \tau$. }
\label{fig:bumpHNL}
\end{figure}

\begin{figure}
\begin{center}
\includegraphics[width=0.7
\columnwidth]{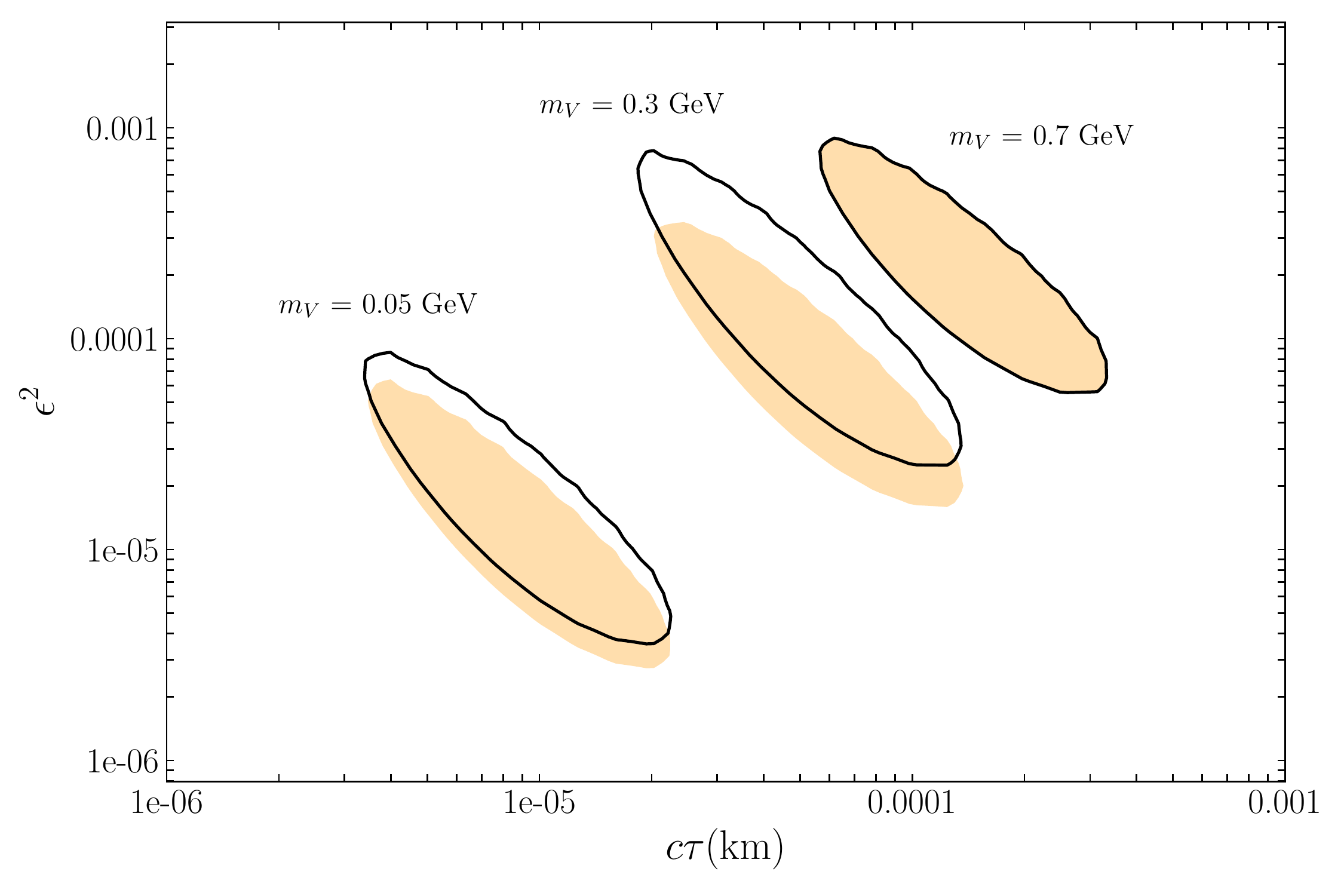} 
\end{center}
\caption{$1\sigma$ best fit regions on dark photons decays in IceCube assuming uncorrelated  $\epsilon^2$ vs $c\tau$. The black lines correspond to considering only the signal from $\pi_0$ decay for $m_V= 0.05$~GeV, only $\eta$ decay for $m_V = 0.3$~GeV and only \bremm for the heavier mass. }
\label{fig:bumpDP}
\end{figure}

Figure~\ref{fig:bumpHNL} shows the 1$\sigma$ regions for 2 d.o.f., \ie the region $\Delta \chi^2 = \chi^2 - \chi^2_{min} \leq 2.3$, where $\chi^2_{min}$ is the global minimum, for a HNL produced from $D, D_s$ and $\tau$ decays. On the other hand, $\chi^2$(no LLP) - $\chi^2_{min} \simeq 7$. Although our results are shown for $m_N=0.5$~GeV, the regions do not change significantly for other values of $m_N$ between 0.5 and 1.5~GeV.
%%%%%%%%%%%%%%%%%%%%%%%%% 
%%%%%%%%%%%%%%%%%%%%%%%%%
For the dark photon scenario, Fig.~\ref{fig:bumpDP} shows the allowed 1$\sigma$ regions (2 d.o.f.) for three values of the dark photon mass. 
%%%%%%%%%%%%%%%%%%%%%%%%%
%%%%%%%%%%%%%%%%%%%%%%%%%
Unfortunately the minimal models discussed in Secs.~\ref{sec:HNL} and~\ref{sec:darkphoton} cannot live on those regions, according to Figs.~\ref{fig:BrvsctauN} (for the HNL) and Fig.~\ref{fig:evsctauV} (for the dark photon).

Regarding the $B-L$ model with HNL, as in the previous cases we get a best fit to IC data away from zero and, in particular, for ${\rm Br}(V\rightarrow N) g_{B-L}^2 \sim 10^{-6}-10^{-5} $ and $c\tau_N\sim 10^{-4}$~km. Unfortunately, the required value of $g_{B-L}$ is excluded by collider constraints~\cite{Ilten:2018crw,Bauer:2018onh,Heeck:2018nzc} in this dark photon mass range and, at the same time, such as small $c\tau_N$ for a HNL at $m_N=0.35$~GeV is at least one order of magnitude too small, assuming that the mixing is below the present constraints from direct searches of HNL~\cite{Atre:2009rg}. 

Before finalizing this section it should be stressed, however, that our IC analysis has been performed using just the energy information available on the data release of Ref.~\cite{Aartsen:2014muf}, and that in the case of decay of an LLP the zenith angle distribution of the signal events could differ significantly from the distributions observed for the data (see Fig.~9 of \cite{Aartsen:2014muf}). More stringent
constraints might be attainable if information on the zenith angle distribution of the events were included in the analysis, which is not possible with the publicly available information.

\section{Conclusions}
\label{sec:conclu}

Very weakly interacting particles beyond the Standard Model (SM) might be light enough to be produced in accelerators, but easy to miss in standard BSM search analyses if they are very long-lived. Many running accelerator experiments, both fixed-target and colliders, are actively exploring new strategies to improve the sensitivity to such exotic signals, and several future experiments are being proposed to target specifically such BSM scenarios. 

 In this paper we have presented the first detailed analysis of the sensitivity of large neutrino detectors, such as Super-Kamiokande and IceCube, to  putative long-lived particles (LLP) that might be produced in atmospheric showers from the decay of standard model mesons and/or bremsstrahlung. We have presented the methodology to evaluate LLP fluxes from these processes, and the procedure to quantify the expected number of signal events. We have considered in particular two minimal and theoretically well-motivated scenarios where the LLP are either heavy neutral leptons produced primarily in meson decays, or dark photons produced from bremsstrahlung or/and radiative $\pi_0$ and $\eta$ decays. We have evaluated bounds using Super-Kamiokande and IceCube data from Refs.~\cite{Aartsen:2014muf,Abe:2017aap}, as a function of the branching ratios (or production rates) of SM $\rightarrow$ LLPs, and the LLP lifetime. The best sensitivity is obtained for lifetimes of ${\mathcal O}(10^{-2} \times {m_{\rm LLP}\over 1{\rm GeV}})$ km in Icecube and ${\mathcal O}(10 \times {m_{\rm LLP}\over 1{\rm GeV}})$ km for Super-Kamiokande. However, in the minimal models considered in this work production and decay are usually strongly correlated  since both are controlled by the same coupling. In this case, we find that atmospheric searches do not lead to competitive bounds. On the other hand,  our bounds might be complementary to other searches in more complex models where production and decay are uncorrelated. As an example for such scenario we considered an extended model with a $B-L$ gauge boson that couples to the HNL. In this case, the production is controlled by the gauge coupling, $g_{B-L}$, while the LLP is the HNL whose lifetime is controlled by its mixing with the light neutrinos.  We have also presented the  sentitivity to this scenario. 
   
Interestingly, in the case of Icecube the addition of an LLP to the SM provides a better fit to the observed data than the prediction obtained in the SM (with a $\Delta \chi^2 \sim 7$), due to a small excess observed at around 30~TeV. This is obtained, for example, for a HNL produced in $D$ meson decays with a ${\rm Br}(D\rightarrow X) \sim 2\cdot 10^{-3}$ and a lifetime $c\tau\sim 10^{-4}$. However, we find that the values of the parameters needed in order to fit the excess lie outside of the allowed regions of parameter space for the three models considered.

Finally, we stress again that no systematic uncertainties have been included in our calculations. A more detailed study, including a realistic implementation of systematic and reconstruction errors by the experimental collaborations, would be mandatory to validate our results. 

\begin{acknowledgments}
We  warmly thank  A.~Fedynitch for useful discussions and clarifications on the usage of the MCEq package, R.~Zukanovich for discussions at the early stages of this work, S.~Palomares-Ruiz for useful comments and discussions, and I.~M.~Shoemaker for comments on the manuscript. PH and CA thank the Fermilab Theory Group for hospitality during the early stages of this work, and PC  thanks the CERN Theory division for support and hospitality during the final stages of this work. The work of VM is funded by CONICYT PFCHA/DOCTORADO BECAS CHILE/2018 - 72180000. CA is supported by NSF grant PHY-1912764. This work was partially supported by grants FPA2017-85985-P, PROMETEO/2019/083, and  the European projects H2020-MSCA-ITN-2015//674896-ELUSIVES and 690575-InvisiblesPlus-H2020-MSCA-RISE-2015. 
\end{acknowledgments}

\appendix
\section{Computation of effective decay areas}
\label{app:decay}

Assuming the fidutial volume of the detector is a cylinder of radius $R$ and height $H$, it is a simple geometrical exercise to obtain $\Delta\ell_{\rm det}$, and the effective area:
\begin{eqnarray}
A^{\rm eff}_{\rm decay}(E_A,\cos\theta) =  |\cos\theta| A_1(E_A, \cos\theta) + |\sin\theta| A_2 (E_A, \cos\theta) ,
\label{eq:aeff}
\end{eqnarray}
where  $A_1$ correspond to the flux entering from the top:
\begin{eqnarray}
A_1(E_A,\cos\theta) = \int_{0}^{R} r dr \int_{0}^{2 \pi} d\phi ~ \left( 1- e^{- { \Delta\ell^{(1)}_{det}( \theta,r,\phi) \over c\tau_{\rm lab}(E_A)}}  \right),
\end{eqnarray}
with 
\begin{eqnarray}
\Delta\ell^{(1)}_{det}( \theta,r,\phi) \equiv  {\rm Min}\left[{H\over |\cos\theta|},{R \sqrt{1- r^2/R^2 \sin^2\phi} + r \cos\phi\over| \sin\theta|} \right],
\end{eqnarray}
and $A_2$ is that entering laterally:
\begin{eqnarray}
A_2(E_A,\cos\theta) = R \int_{0}^{H}dx \int_{-{\pi\over 2}}^{{\pi\over 2}} d\phi ~ \left( 1- e^{-  { \Delta\ell^{(2)}_{det}( \theta,x,\phi) \over  c\tau_{\rm lab}(E_A)}}  \right),
\end{eqnarray}
with 
\begin{eqnarray}
\Delta\ell^{(2)}_{det}( \theta,x,\phi) \equiv  {\rm Min}\left[{H-x \over |\cos\theta|}, {2 R \cos\phi\over |\sin\theta|} \right].
\end{eqnarray}

\section{Computation of parent particle fluxes in the atmosphere}
\label{app:fluxes}

In all our calculations, the SM parent particle fluxes in the atmosphere have been computed using the MCEq software~\cite{Fedynitch:2015zma,Fedynitch:2012fs}, with the SYBILL-2.3 hadronic interaction model~\cite{Fedynitch:2018cbl}, the Hillas-Gaisser cosmic-ray model~\cite{Gaisser:2011cc} and the NRLMSISE-00 atmospheric model~\cite{Picone:2002go}. Obtaining the parent particle fluxes is however not straightforward, because only the flux for protons and unstable particles that are relatively long-lived compared to their interaction length can be directly extracted using MCEq. A possible strategy to circumvent this is to manually switch off the decay of the parent particle and extract the flux, which then needs to be corrected to account for the decay\footnote{We thank A. Fedynitch for providing guidance on this point.}. 

To simplify notation, we denote the differential flux of the parent particle $P$  by $\phi_P$:
\begin{eqnarray}
\phi_P \equiv {d \Phi_P\over d E }.
\end{eqnarray}

Let us assume that ${\tilde\phi}_P$ is the differential flux of parent particle $P$ \emph{assuming} that $P$ is stable\footnote{In MCEq, this can be easily extracted by manually switching off the decay of the particle $P$ (using the advanced feature settings). }. Assuming that $P$ is directly produced in nucleon interactions, this flux satisfies a cascade equation of the form~\cite{Gondolo:1995fq}
\begin{eqnarray}
\label{eq:cascade}
{d\tilde\phi_P\over dX} = -{\tilde\phi_P \over \lambda_P} +  Z_{NP} {\phi_N\over \lambda_N}+Z_{PP} {\tilde\phi_P\over \lambda_P},
\end{eqnarray}
where $\phi_N$ is the flux of nucleons, and the spectrum-weighted momenta are defined as 
\begin{eqnarray}
Z_{kh}= \int_E^\infty ~dE_P {\phi_k(E_P)\over \phi_k(E)} {\lambda_k(E) \over \lambda_k(E_P)} { d n(k \mathcal{N} \rightarrow h Y; E_P, E)\over dE} \, .
\end{eqnarray}
Here, $\lambda_k$ is the particle interaction length of hadron $k$ and ${dn(k \mathcal{N} \rightarrow h Y)\over dE}$  is the number of  hadrons $h$ produced with energy between $E$ and $E+dE$, in the scattering of hadron $k$ with energy $E_P$ on a nucleus $\mathcal{N}$. 

The second and third terms in Eq.~\ref{eq:cascade} are the production and regeneration terms, respectively, while the first term is the absorption term. Had the particle decayed, the correct cascade equation should have also included a decay term:
\begin{eqnarray}
{d\phi_P\over dX} = -{\phi_P \over \rho d_P}-{\phi_P \over \lambda_P}  +  Z_{NP} {\phi_N\over \lambda_N}+Z_{PP} {\phi_P\over \lambda_P},
\label{eq:gondolo}
\end{eqnarray}
where $d_P$ is the decay length of the parent particle in the laboratory frame and $\rho$ is the column density at the point with slant depth $X$. 
It is easy to see that from $\tilde\phi_P$  we can  get  the solution to Eq.~(\ref{eq:gondolo}) as
\begin{eqnarray}
 \phi_P = \int_0^X  ~dX' ~e^{-(X-X')\left({1\over \rho d_P} + {1 \over \Lambda_P}\right)}  \left({d \tilde\phi_P  (X') \over dX}+ {\tilde\phi_P(X') \over \Lambda_P}  \right),
\end{eqnarray}
with 
\begin{eqnarray}
\Lambda_P \equiv {\lambda_P\over 1-Z_{PP}}.
\end{eqnarray}
To the best of our knowledge, although MCEq allows to obtain the values of $\lambda_P$ it does not allow to extract the values of $Z_{PP}$ directly. Therefore, we assume $Z_{\tau\tau}=0$ while for the rest of the mesons considered we assume Feynmann scaling (that is, $Z_{MM}=0.3$) since, according to Fig.~4 in Ref.~\cite{Gondolo:1995fq}, this is a reasonable approximation for charged pions and kaons. In any case, we do not expect our results to change significantly if these assumptions are modified.

With this procedure we get a reasonable agreement with the output of MCEq available at large energies, where decay can be ignored. We do a small rescaling of our approximate result by matching the two curves in the overlapping region. 

Finally, in the case of $D$ and $D_s$ mesons both the interaction and regeneration terms in Eq.~\ref{eq:gondolo} can be neglected. In this case, the meson flux may be computed directly from the flux of protons in the atmosphere, following Ref.~\cite{Gondolo:1995fq} (Eqs.~(25-27) in that reference):
\begin{equation}
\label{eq:gondolo-approx}
\phi_M^{low} \simeq Z_{NM}\frac{\rho d_M}{\lambda_N} \phi_N(X), \, .
\end{equation}
where $\phi_N(X)$ and the yields, $Z_{NM}$, are extracted from MCEq.
We have also checked that this method gives a good agreement with our fluxes, in the case of $D$ and $D_s$ mesons.

\bibliographystyle{JHEP}
\bibliography{biblio}

\end{document}